\pdfoutput=1
\documentclass[12pt,a4paper]{article}

\usepackage{ifthen} % for conditional statements
\newboolean{pdflatex}
\setboolean{pdflatex}{true} % False for eps figures 

\newboolean{articletitles}
\setboolean{articletitles}{true} % False removes titles in references

\newboolean{uprightparticles}
\setboolean{uprightparticles}{false} %True for upright particle symbols

\newboolean{inbibliography}
\setboolean{inbibliography}{false} %True once you enter the bibliography

\def\phani{\phantom{1}}

\def\phanii{\phantom{11}}

\usepackage{microtype}
\usepackage{lineno}  % for line numbering during review
\usepackage{xspace} % To avoid problems with missing or double spaces after
\usepackage{caption} %these three command get the figure and table captions automatically small

\usepackage{graphicx}  % to include figures (can also use other packages)
\usepackage{rotating}
\usepackage{color}
\usepackage{colortbl}
\usepackage{amsmath} % Adds a large collection of math symbols
\usepackage{amssymb}
\usepackage{amsfonts}
\usepackage{upgreek} % Adds in support for greek letters in roman typeset

\newcommand*\patchAmsMathEnvironmentForLineno[1]{%
\expandafter\let\csname old#1\expandafter\endcsname\csname #1\endcsname
\expandafter\let\csname oldend#1\expandafter\endcsname\csname
end#1\endcsname
 \renewenvironment{#1}%
   {\linenomath\csname old#1\endcsname}%
   {\csname oldend#1\endcsname\endlinenomath}%
}
\newcommand*\patchBothAmsMathEnvironmentsForLineno[1]{%
  \patchAmsMathEnvironmentForLineno{#1}%
  \patchAmsMathEnvironmentForLineno{#1*}%
}
\AtBeginDocument{%
\patchBothAmsMathEnvironmentsForLineno{equation}%
\patchBothAmsMathEnvironmentsForLineno{align}%
\patchBothAmsMathEnvironmentsForLineno{flalign}%
\patchBothAmsMathEnvironmentsForLineno{alignat}%
\patchBothAmsMathEnvironmentsForLineno{gather}%
\patchBothAmsMathEnvironmentsForLineno{multline}%
\patchBothAmsMathEnvironmentsForLineno{eqnarray}%
}

\usepackage{hyperref}    % Hyperlinks in references
\usepackage[all]{hypcap} % Internal hyperlinks to floats.

\def\WSdkpi {\ensuremath{\Dp\Kp\pim}\xspace}
\def\RSdkpi {\ensuremath{\Dm\Kp\pip}\xspace}
\def\BtoWSdkpi {\ensuremath{\Bp\to\WSdkpi}\xspace}
\def\BtoRSdkpi {\ensuremath{\Bp\to\RSdkpi}\xspace}

\def\BtoRSdpipi {\ensuremath{\Bp\to\Dm\pip\pip}\xspace}
\def\BtoRSdstarpipi {\ensuremath{\Bp\to\Dstarm\pip\pip}\xspace}
\def\BtoRSdstarkpi {\ensuremath{\Bp\to\Dstarm\Kp\pip}\xspace}

\def\Btodpipi {\ensuremath{\BtoRSdpipi}\xspace}
\def\Btodstarpipi {\ensuremath{\BtoRSdstarpipi}\xspace}
\def\Btodstarkpi {\ensuremath{\BtoRSdstarkpi}\xspace}
\def\Btodskpi {\ensuremath{\Bp\to\Dsm\Kp\pip}\xspace}
\def\Btodanddstarpipi {\ensuremath{\Bp\to D^{(*)-}\pip\pip}\xspace}

\def\BtoDstarK {\ensuremath{\Bp\to D^{*}_{2}(2460)^{0}\Kp}\xspace}
\def\BtoKstarD {\ensuremath{\Bp\to \Dp K^{*}(892)^{0}}\xspace}

\def\Bstodkpipi {\ensuremath{\Bsb\to\Dp\Kp\pim\pim}\xspace}

\def\lhcb {\mbox{LHCb}\xspace}

\def\babar  {\mbox{BaBar}\xspace}

\def\MagUp {\mbox{\em Mag\kern -0.05em Up}\xspace}

\ifthenelse{\boolean{uprightparticles}}%
{

 \def\Ppi         {\ensuremath{\uppi}\xspace}

 \def\PDelta      {\ensuremath{\Delta}\xspace}                 
 \def\PXi      {\ensuremath{\Xi}\xspace}                 
 \def\PLambda      {\ensuremath{\Lambda}\xspace}                 
 \def\PSigma      {\ensuremath{\Sigma}\xspace}                 
 \def\POmega      {\ensuremath{\Omega}\xspace}                 
 \def\PUpsilon      {\ensuremath{\Upsilon}\xspace}                 
 
 \def\PB      {\ensuremath{\mathrm{B}}\xspace}                 
                  
 \def\PD      {\ensuremath{\mathrm{D}}\xspace}

 \def\PK      {\ensuremath{\mathrm{K}}\xspace}

 \def\Pb      {\ensuremath{\mathrm{b}}\xspace}                 
 \def\Pc      {\ensuremath{\mathrm{c}}\xspace}

 \def\Pi      {\ensuremath{\mathrm{i}}\xspace}

 \def\Ps      {\ensuremath{\mathrm{s}}\xspace}                 
                  
 \def\Pu      {\ensuremath{\mathrm{u}}\xspace}

}
{

 \def\Ppi         {\ensuremath{\pi}\xspace}

 \mathchardef\PDelta="7101
 \mathchardef\PXi="7104
 \mathchardef\PLambda="7103
 \mathchardef\PSigma="7106
 \mathchardef\POmega="710A
 \mathchardef\PUpsilon="7107
                  
 \def\PB      {\ensuremath{B}\xspace}                 
                  
 \def\PD      {\ensuremath{D}\xspace}

 \def\PK      {\ensuremath{K}\xspace}

 \def\Pb      {\ensuremath{b}\xspace}                 
 \def\Pc      {\ensuremath{c}\xspace}

 \def\Pi      {\ensuremath{i}\xspace}

 \def\Ps      {\ensuremath{s}\xspace}                 
                  
 \def\Pu      {\ensuremath{u}\xspace}

}

\makeatletter
\ifcase \@ptsize \relax% 10pt
  \newcommand{\miniscule}{\@setfontsize\miniscule{4}{5}}% \tiny: 5/6
\or% 11pt
  \newcommand{\miniscule}{\@setfontsize\miniscule{5}{6}}% \tiny: 6/7
\or% 12pt
  \newcommand{\miniscule}{\@setfontsize\miniscule{5}{6}}% \tiny: 6/7
\fi
\makeatother

\DeclareRobustCommand{\optbar}[1]{\shortstack{{\miniscule (\rule[.5ex]{1.25em}{.18mm})}
  \\ [-.7ex] $#1$}}

\def\uquark    {{\ensuremath{\Pu}}\xspace}
\def\uquarkbar {{\ensuremath{\overline \uquark}}\xspace}

\def\squark    {{\ensuremath{\Ps}}\xspace}

\def\cquark    {{\ensuremath{\Pc}}\xspace}
\def\cquarkbar {{\ensuremath{\overline \cquark}}\xspace}

\def\bquark    {{\ensuremath{\Pb}}\xspace}
\def\bquarkbar {{\ensuremath{\overline \bquark}}\xspace}

\def\pion   {{\ensuremath{\Ppi}}\xspace}
\def\piz    {{\ensuremath{\pion^0}}\xspace}

\def\pip    {{\ensuremath{\pion^+}}\xspace}
\def\pim    {{\ensuremath{\pion^-}}\xspace}
\def\pipm   {{\ensuremath{\pion^\pm}}\xspace}
\def\pimp   {{\ensuremath{\pion^\mp}}\xspace}

\def\kaon    {{\ensuremath{\PK}}\xspace}
  \def\Kbar    {{\kern 0.2em\overline{\kern -0.2em \PK}{}}\xspace}

\def\KorKbar    {\kern 0.18em\optbar{\kern -0.18em K}{}\xspace}

\def\Kp      {{\ensuremath{\kaon^+}}\xspace}
\def\Km      {{\ensuremath{\kaon^-}}\xspace}
\def\Kpm     {{\ensuremath{\kaon^\pm}}\xspace}
\def\Kmp     {{\ensuremath{\kaon^\mp}}\xspace}
\def\KS      {{\ensuremath{\kaon^0_{\rm\scriptscriptstyle S}}}\xspace}

\def\Kstar   {{\ensuremath{\kaon^*}}\xspace}

  \def\Dbar    {{\kern 0.2em\overline{\kern -0.2em \PD}{}}\xspace}
\def\D       {{\ensuremath{\PD}}\xspace}

\def\DorDbar    {\kern 0.18em\optbar{\kern -0.18em D}{}\xspace}
\def\Dz      {{\ensuremath{\D^0}}\xspace}
\def\Dzb     {{\ensuremath{\Dbar{}^0}}\xspace}
\def\Dp      {{\ensuremath{\D^+}}\xspace}
\def\Dm      {{\ensuremath{\D^-}}\xspace}
\def\Dpm     {{\ensuremath{\D^\pm}}\xspace}

\def\Dstarm  {{\ensuremath{\D^{*-}}}\xspace}

\def\Dsm     {{\ensuremath{\D^-_\squark}}\xspace}

\def\B       {{\ensuremath{\PB}}\xspace}
\def\Bbar    {{\ensuremath{\kern 0.18em\overline{\kern -0.18em \PB}{}}}\xspace}

\def\BorBbar    {\kern 0.18em\optbar{\kern -0.18em B}{}\xspace}
\def\Bz      {{\ensuremath{\B^0}}\xspace}

\def\Bu      {{\ensuremath{\B^+}}\xspace}
\def\Bub     {{\ensuremath{\B^-}}\xspace}
\def\Bp      {{\ensuremath{\Bu}}\xspace}
\def\Bm      {{\ensuremath{\Bub}}\xspace}
\def\Bpm     {{\ensuremath{\B^\pm}}\xspace}

\def\Bd      {{\ensuremath{\B^0}}\xspace}
\def\Bs      {{\ensuremath{\B^0_\squark}}\xspace}
\def\Bsb     {{\ensuremath{\Bbar{}^0_\squark}}\xspace}

  \def\Y#1S{\ensuremath{\PUpsilon{(#1S)}}\xspace}% no space before {...}!

\def\Lbar        {{\ensuremath{\kern 0.1em\overline{\kern -0.1em\PLambda}}}\xspace}
\def\LorLbar    {\kern 0.18em\optbar{\kern -0.18em \PLambda}{}\xspace}

\def\BF         {{\ensuremath{\cal B}}\xspace}

\def\BR         {\BF}
\newcommand{\decay}[2]{\ensuremath{#1\!\to #2}\xspace}         % {\Pa}{\Pb \Pc}

\def\to                 {\ensuremath{\rightarrow}\xspace}

\def\CP                {{\ensuremath{C\!P}}\xspace}

\def\AT#1     {\ensuremath{A_{\mathrm{T}}^{#1}}\xspace}           % 2

\def\C#1      {\ensuremath{\mathcal{C}_{#1}}\xspace}                       % 9
\def\Cp#1     {\ensuremath{\mathcal{C}_{#1}^{'}}\xspace}                    % 7
\def\Ceff#1   {\ensuremath{\mathcal{C}_{#1}^{\mathrm{(eff)}}}\xspace}        % 9  
\def\Cpeff#1  {\ensuremath{\mathcal{C}_{#1}^{'\mathrm{(eff)}}}\xspace}       % 7
\def\Ope#1    {\ensuremath{\mathcal{O}_{#1}}\xspace}                       % 2
\def\Opep#1   {\ensuremath{\mathcal{O}_{#1}^{'}}\xspace}                    % 7

\newcommand{\tev}{\ifthenelse{\boolean{inbibliography}}{\ensuremath{~T\kern -0.05em eV}\xspace}{\ensuremath{\mathrm{\,Te\kern -0.1em V}}}\xspace}
\newcommand{\gev}{\ensuremath{\mathrm{\,Ge\kern -0.1em V}}\xspace}
\newcommand{\mev}{\ensuremath{\mathrm{\,Me\kern -0.1em V}}\xspace}
\newcommand{\kev}{\ensuremath{\mathrm{\,ke\kern -0.1em V}}\xspace}
\newcommand{\gevnsp}{\ensuremath{\mathrm{Ge\kern -0.1em V}}\xspace}
\newcommand{\mevnsp}{\ensuremath{\mathrm{Me\kern -0.1em V}}\xspace}
\newcommand{\kevnsp}{\ensuremath{\mathrm{ke\kern -0.1em V}}\xspace}
\newcommand{\ev}{\ensuremath{\mathrm{\,e\kern -0.1em V}}\xspace}
\newcommand{\gevc}{\ensuremath{{\mathrm{\,Ge\kern -0.1em V\!/}c}}\xspace}
\newcommand{\mevc}{\ensuremath{{\mathrm{\,Me\kern -0.1em V\!/}c}}\xspace}
\newcommand{\gevcc}{\ensuremath{{\mathrm{\,Ge\kern -0.1em V\!/}c^2}}\xspace}
\newcommand{\gevgevcccc}{\ensuremath{{\mathrm{\,Ge\kern -0.1em V^2\!/}c^4}}\xspace}
\newcommand{\mevcc}{\ensuremath{{\mathrm{\,Me\kern -0.1em V\!/}c^2}}\xspace}

\def\invfb   {\ensuremath{\mbox{\,fb}^{-1}}\xspace}

\newcommand{\stat}{\ensuremath{\mathrm{\,(stat)}}\xspace}
\newcommand{\syst}{\ensuremath{\mathrm{\,(syst)}}\xspace}

\newcommand{\chisq}{\ensuremath{\chi^2}\xspace}

\def\gsim{{~\raise.15em\hbox{$>$}\kern-.85em
          \lower.35em\hbox{$\sim$}~}\xspace}
\def\lsim{{~\raise.15em\hbox{$<$}\kern-.85em
          \lower.35em\hbox{$\sim$}~}\xspace}

\def\sPlot{\mbox{\em sPlot}\xspace}
\def\tell1  {TELL1\xspace}
\def\ukl1   {UKL1\xspace}

\usepackage{cite} % Allows for ranges in citations
\usepackage{mciteplus}
\usepackage{multirow}

\begin{document}

%%%%%%%%%%%%%%%%%%%%%%%%%
%%%%% Title     %%%%%%%%%
%%%%%%%%%%%%%%%%%%%%%%%%%
\renewcommand{\thefootnote}{\fnsymbol{footnote}}
\setcounter{footnote}{1}

\begin{titlepage}
\pagenumbering{roman}

% Header ---------------------------------------------------
\vspace*{-1.5cm}
\centerline{\large EUROPEAN ORGANIZATION FOR NUCLEAR RESEARCH (CERN)}
\vspace*{1.5cm}
\hspace*{-0.5cm}
\begin{tabular*}{\linewidth}{lc@{\extracolsep{\fill}}r}
\ifthenelse{\boolean{pdflatex}}% Logo format choice
{\vspace*{-2.7cm}\mbox{\!\!\!\includegraphics[width=.14\textwidth]{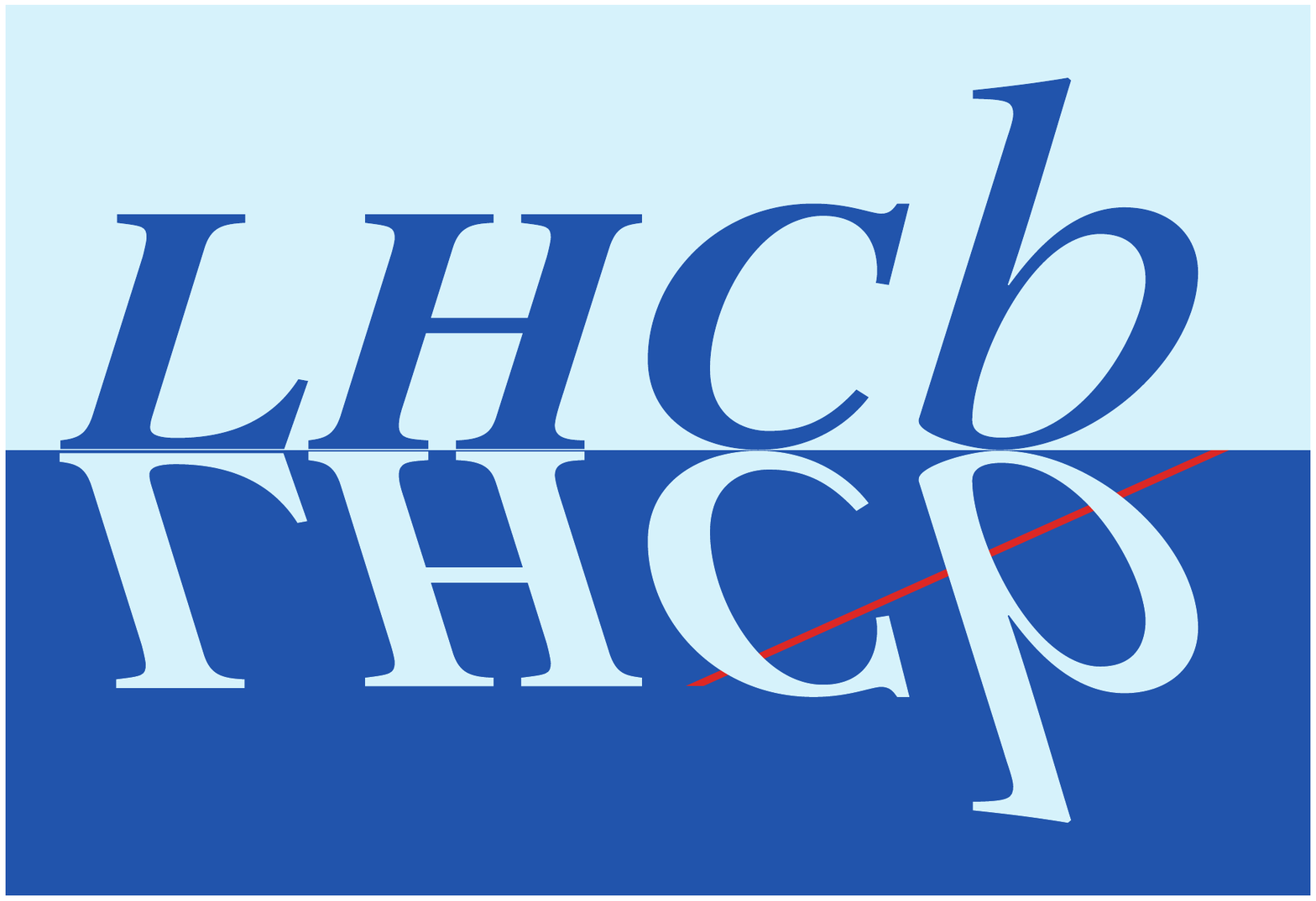}} & &}%
{\vspace*{-1.2cm}\mbox{\!\!\!\includegraphics[width=.12\textwidth]{lhcb-logo.eps}} & &}%
\\
 & & CERN-PH-EP-2015-313 \\  % ID 
 & & LHCb-PAPER-2015-054 \\  % ID 
 & & 21 June 2016 \\ % Date - Can also hardwire e.g.: 23 March 2010
 & & \\
% not in paper \hline
\end{tabular*}

\vspace*{2.5cm}

% Title --------------------------------------------------
{\bf\boldmath\huge
\begin{center}
  First observation of the rare $\BtoWSdkpi$ decay
\end{center}
}

\vspace*{1.5cm}

% Authors -------------------------------------------------
\begin{center}
%In the footnote, replace 'paper' by 'letter' in case of submission to PRL or PLB 
The LHCb collaboration\footnote{Authors are listed at the end of this paper.}
\end{center}

\vspace{\fill}

% Abstract -----------------------------------------------
\begin{abstract}
\noindent
The $\BtoWSdkpi$ decay is observed in a data sample corresponding to $3.0\invfb$ of $pp$ collision data recorded by the LHCb experiment during 2011 and 2012.
The signal significance is $8\,\sigma$ and the branching fraction is measured to be
$\BR\left(\BtoWSdkpi\right) = (5.31 \pm 0.90 \pm 0.48 \pm 0.35)\times 10^{-6}$, where the uncertainties are statistical, systematic and due to the normalisation mode $\BtoRSdkpi$, respectively.
The Dalitz plot appears to be dominated by broad structures.
Angular distributions are exploited to search for quasi-two-body contributions from $\BtoDstarK$ and $\BtoKstarD$ decays. No significant signals are observed and upper limits are set on their branching fractions.
\end{abstract}

\vspace*{1.5cm}

\begin{center}
  Published in Phys.~Rev.~D.~(R)
\end{center}

\vspace{\fill}

{\footnotesize 
\centerline{\copyright~CERN on behalf of the \lhcb collaboration, licence \href{http://creativecommons.org/licenses/by/4.0/}{CC-BY-4.0}.}}
\vspace*{2mm}

\end{titlepage}

%%%%%%%%%%%%%%%%%%%%%%%%%%%%%%%%
%%%%%  EOD OF TITLE PAGE  %%%%%%
%%%%%%%%%%%%%%%%%%%%%%%%%%%%%%%%

%  empty page follows the title page ----
\newpage
\setcounter{page}{2}
\mbox{~}

\cleardoublepage

\renewcommand{\thefootnote}{\arabic{footnote}}
\setcounter{footnote}{0}
\pagestyle{plain} % restore page numbers for the main text
\setcounter{page}{1}
\pagenumbering{arabic}

%% Uncomment during review phase. 
%% Comment before a final submission.
%\linenumbers

A key goal of flavour physics is to determine precisely the angle $\gamma$ of the unitarity triangle constructed from pairs of elements of the Cabibbo-Kobayashi-Maskawa (CKM) quark mixing matrix~\cite{Cabibbo:1963yz,Kobayashi:1973fv}.
The value of $\gamma \equiv \arg\left[-V_{ud}^{}V_{ub}^*/(V_{cd}^{}V_{cb}^*)\right]$ is currently known to a precision of only about $10^\circ$~\cite{Bona:2005vz,Charles:2004jd,LHCb-CONF-2014-004}, which limits the sensitivity of tests of the Standard Model through global fits to the CKM matrix parameters.

\begin{figure}[!b]
  \centering
  \includegraphics[width=0.33\textwidth]{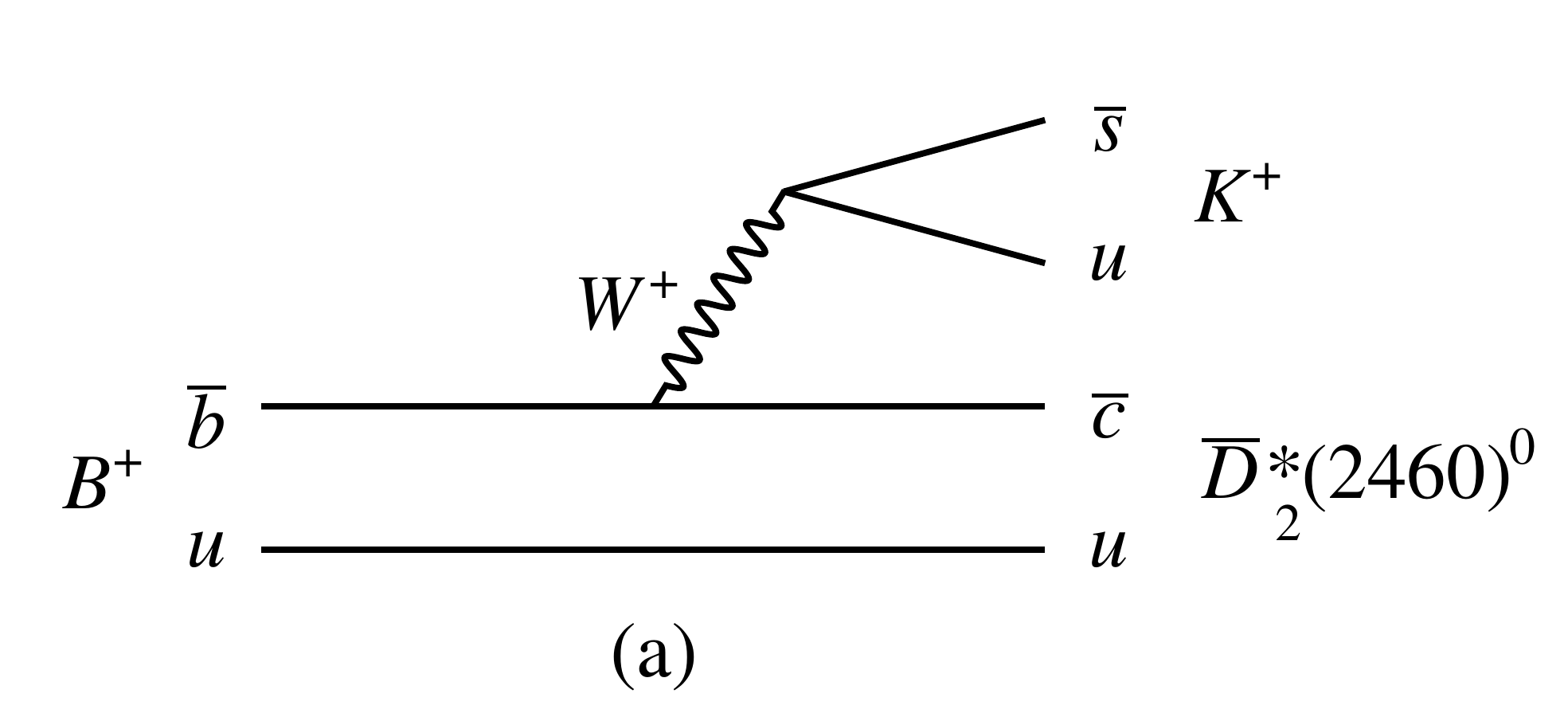}
  \hspace{-2mm}
  \includegraphics[width=0.33\textwidth]{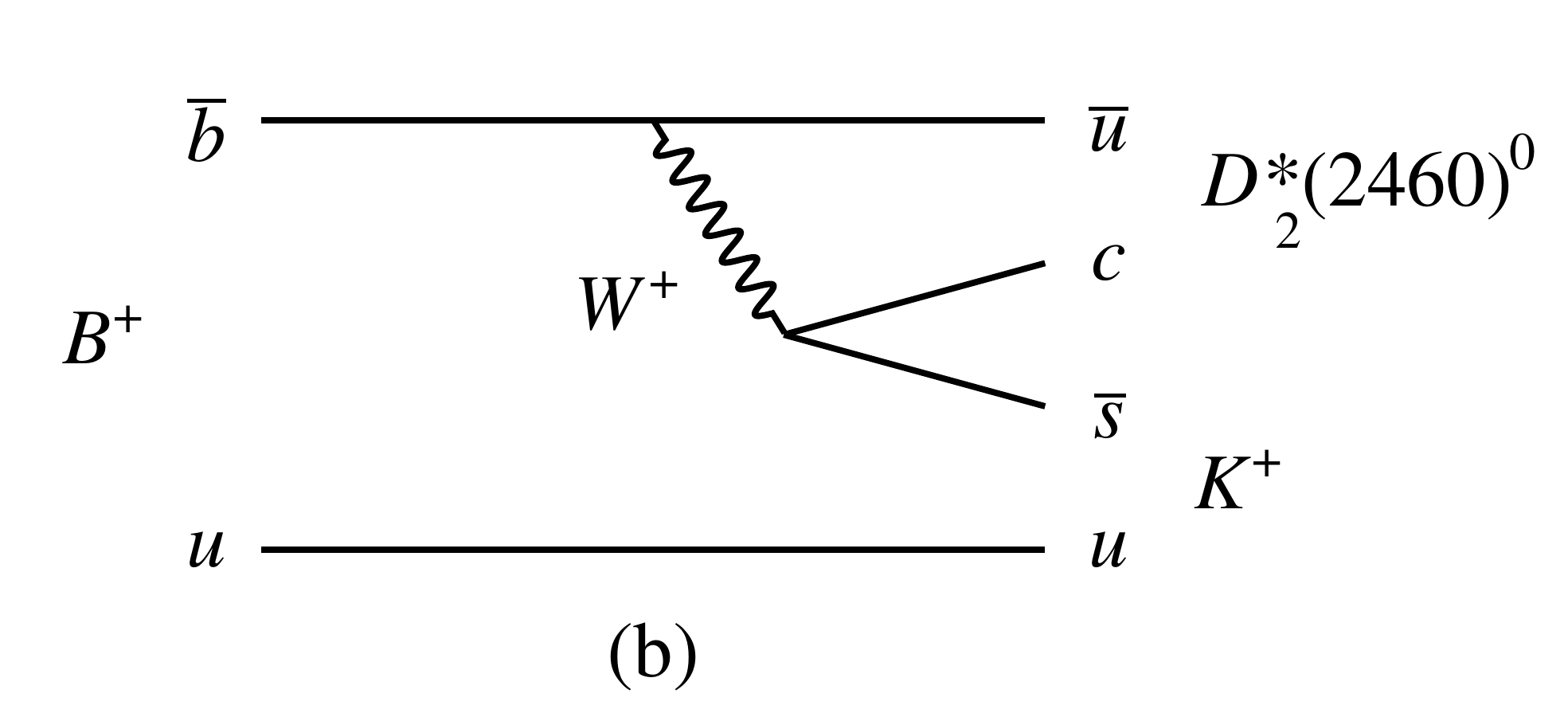}
  \hspace{-2mm}
  \includegraphics[width=0.33\textwidth]{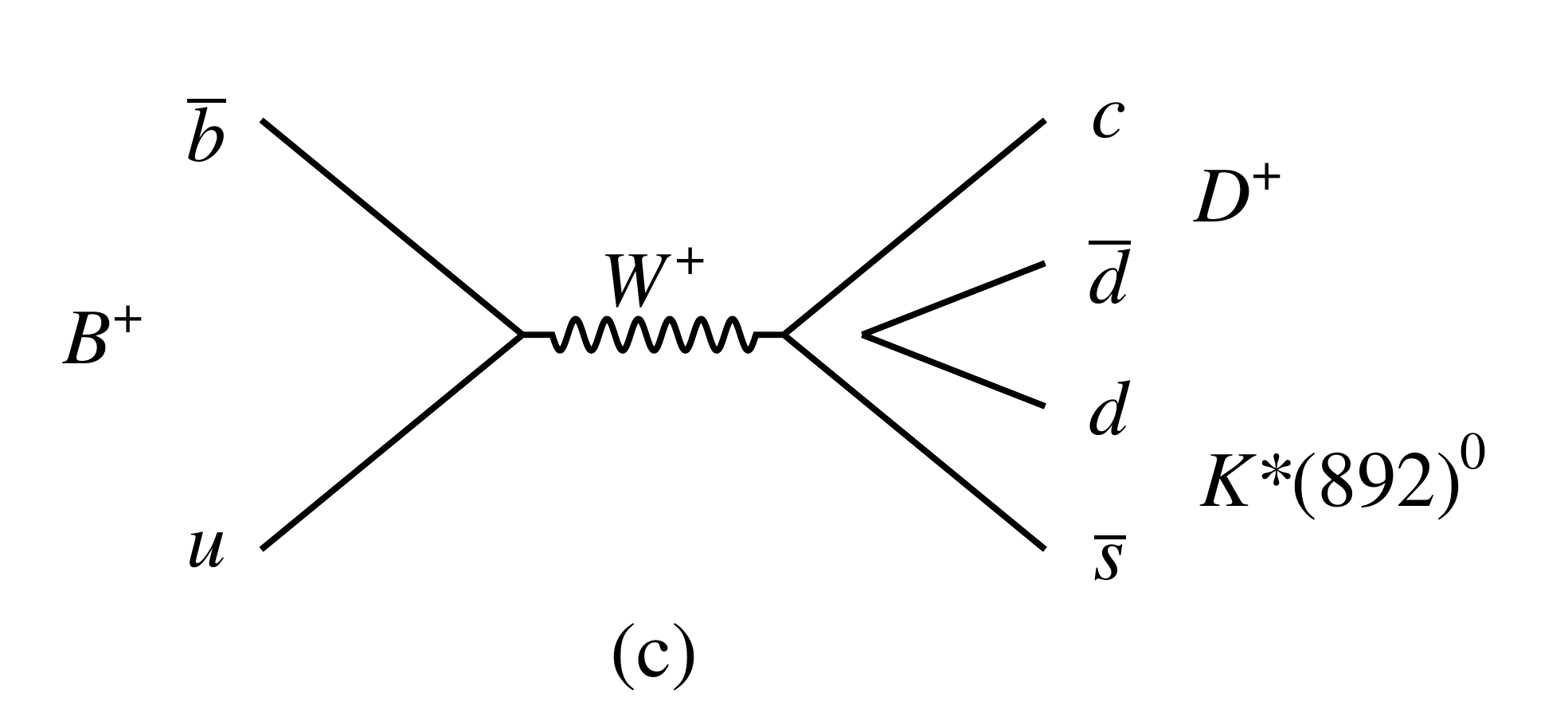}
  \caption{\small
    Decay diagrams for (a)~$\decay{\Bp}{\Dbar{}^*_2(2460)^0 K^+}$, (b)~$\decay{\Bp}{D^*_2(2460)^0 K^+}$ and (c)~\mbox{$\Bp \to \Dp K^*(892)^0$} channels.
  }
  \label{fig:feynman}
\end{figure}

A powerful method to determine $\gamma$, known as the GLW method~\cite{Gronau:1990ra,Gronau:1991dp}, is to use $\Bp \to D \Kp$ decays with the neutral $D$ meson decaying to \CP eigenstates. 
The $\bquarkbar \to \cquarkbar$ and $\bquarkbar \to \uquarkbar$ amplitudes both contribute to the decay, and the sensitivity to $\gamma$ comes from their interference.
A challenge with this method is that the ratio of magnitudes of the suppressed and favoured amplitudes, $r_B$, is not known independently and must be determined simultaneously with $\gamma$.
This is usually addressed by using in addition other decays of the $\D$ meson that provide complementary information on $r_B$ and $\gamma$~\cite{Atwood:1996ci,Atwood:2000ck}.

In the case of $\Bp \to D^{**} \Kp$ decays, where $D^{**}$ represents an excited $\D$ or $\Dbar$ meson such as the $D_2^*(2460)$ state which can decay to both $\Dpm\pimp$ and $\D\piz$, it is possible to obtain a clean determination of $r_B$~\cite{Sinha:2004ct}.
The relative branching fractions of the $\bquarkbar \to \uquarkbar$ mediated $\Bp \to D^{**0} \Kp \to \Dp \pim \Kp$ and the $\bquarkbar \to \cquarkbar$ mediated $\Bp \to \Dbar{}^{**0} \Kp \to \Dm \pip \Kp$ processes give the value of $r^{2}_B$, whilst the $\Bp \to D^{**} \Kp \to D \piz \Kp$ final state, where the $D$ meson is reconstructed using \CP eigenstate decay modes, provides sensitivity to $\gamma$.
Decay diagrams for $\decay{\Bp}{\Dbar{}^*_2(2460)^0 K^+}$ and $\decay{\Bp}{D^*_2(2460)^0 K^+}$ decays are shown in Fig.~\ref{fig:feynman}(a) and~(b). 

Knowledge of the resonant structure of $\BtoWSdkpi$ and $\BtoRSdkpi$ decays is therefore needed. The latter channel has recently been studied with a Dalitz plot analysis~\cite{LHCb-PAPER-2015-007}.
Such a study would be difficult with the low yields expected for $\BtoWSdkpi$ decays in the available data samples, but an alternative approach exploiting the angular decay information to separate different spin states is viable in the region of the narrow $D_2^*(2460)^0$ resonance.  
The same method can also be used to search for $\Bp \to \Dp K^*(892)^0$ decays, which contribute to the $\Dp\Kp\pim$ final state and are of interest since they are mediated by annihilation amplitudes, as shown in Fig.~\ref{fig:feynman}(c).
A previous LHCb analysis of this mode 
set an upper limit ${\cal B}(\Bp \to \Dp K^*(892)^0) < 1.8 \times 10^{-6}$ at the 90\,\% credibility level~\cite{LHCb-PAPER-2012-025}.

In this paper, the $\BtoWSdkpi$ channel is studied for the first time, and searches for $\Bp \to D_2^*(2460)^0\Kp$ and $\Bp \to \Dp K^*(892)^0$ decays are performed.
The $\Dp$ meson is reconstructed in the $\Km\pip\pip$ final state. 
(The inclusion of charge conjugate processes is implied.)
The $\BtoRSdkpi$ decay~\cite{LHCb-PAPER-2015-007} is used for normalisation. 
The analysis is based on $3.0\invfb$ of $pp$ collision data collected with the LHCb detector during 2011 and 2012.
The \lhcb detector is a single-arm forward spectrometer covering the \mbox{pseudorapidity} range $2<\eta <5$, described in detail in Refs.~\cite{Alves:2008zz,LHCb-DP-2014-002}.
Simulated events are produced using the software described in Refs.~\cite{Sjostrand:2007gs,*Sjostrand:2006za,LHCb-PROC-2010-056,Lange:2001uf,Agostinelli:2002hh,*Allison:2006ve,Golonka:2005pn,LHCb-PROC-2011-006}.
To reduce the risk of biasing results, all analysis procedures were established before the data in the signal region were examined.

Candidates consistent with the decay chains $\BtoWSdkpi$ and $\BtoRSdkpi$ with $\Dpm\to\Kmp\pipm\pipm$ are selected.
The criteria for $\BtoRSdkpi$ and $\BtoWSdkpi$ candidates are identical, except for charge requirements, and are very similar to those described in Ref.~\cite{LHCb-PAPER-2015-007}.
A loose preselection is applied before two neural network classifiers (NNs)~\cite{Feindt:2006pm} are used to separate signal decays from background events. 
The first NN separates true $\Dpm\to\Kmp\pipm\pipm$ decays from random combinations, and the second (NN2) identifies signal $\Bp$ decays. 
Both NNs are trained with a sample of candidates from the topologically similar $\Btodpipi$ decay.
Additional selection requirements are imposed to reject contributions from specific decay modes.
Only candidates with $DK$ mass, $m(DK)$, less than $5140\mevcc$ are kept, in order to remove backgrounds from $\Bd\to\Dm\Kp$ decays combined with a random pion candidate.
Similarly, potential $\Bd\to\Dm\pip$ background is eliminated by requiring $m(D\pi) < 4790\mevcc$.
Contributions from $\Bp\to\Dp\Dzb$ decays, with $\Dzb \to \Kp\pim$, are removed by rejecting candidates within $\sim3\,\sigma_{K\pi}$ of the \Dzb mass, where $\sigma_{K\pi}$ is the $\Kp\pim$ mass resolution, corresponding to $1830< m(K\pi)<1890\mevcc$.
Although each of these backgrounds affects only one of the final states, the vetoes are applied to both to avoid biasing the relative efficiency.

Signal candidates with invariant mass in the range $5100$--$5800\mevcc$ are retained for further analysis.
Following all selection requirements, fewer than $1\,\%$ of events contain more than one candidate; all are retained.

% In order to optimise the sensitivity, a loose requirement is placed on the output of NN2.
Extended maximum likelihood fits to the distributions of candidates in NN2 output and in \B candidate mass are used to determine the yields of $\BtoWSdkpi$ and $\BtoRSdkpi$ decays.
Similar fitting techniques have been used successfully in several previous LHCb analyses~\cite{LHCb-PAPER-2013-046,LHCb-PAPER-2014-052,LHCb-PAPER-2015-012,LHCb-PAPER-2015-048}.
A loose requirement is placed on the output of NN2 and the remaining data in each sample are divided into six bins of the NN2 output variable, each containing a similar number of signal decays.
This binning scheme enhances the sensitivity whilst giving stable fit performance.

The \B candidate mass shapes in the fit to $\BtoRSdkpi$ candidates are modelled in the same way as described in Ref.~\cite{LHCb-PAPER-2015-007}.
The signal is described by the sum of two Crystal Ball (CB)~\cite{Skwarnicki:1986xj} functions, with a common mean. 
The tails on both sides of the peak are described by parameters that are fixed to values found in fits to simulated samples. 
Components are included for combinatorial background, modelled with an exponential shape, and for partially reconstructed backgrounds from $\Btodstarkpi$ decays and misidentified $\Btodanddstarpipi$ and $\Btodskpi$ decays, for which non-parametric descriptions are determined from simulation.
Data-driven estimates of the misidentification probabilities,  
the phase-space distributions of the $\Btodanddstarpipi$ decays~\cite{Abe:2003zm,Aubert:2009wg} and the relative branching fractions of the $\Btodpipi$ and $\Btodstarpipi$ modes~\cite{Abe:2003zm,PDG2014}, are used to obtain these shapes.
For signal and partially reconstructed and combinatorial background components, the relative yields in each NN2 bin are free parameters of the fit, while those of misidentified $\Btodanddstarpipi$ and $\Btodskpi$ decays are taken to be the same as for signal decays, since their NN2 responses are expected to be very similar.

\begin{figure}[!tb]
\centering
\includegraphics[width=0.49\textwidth]{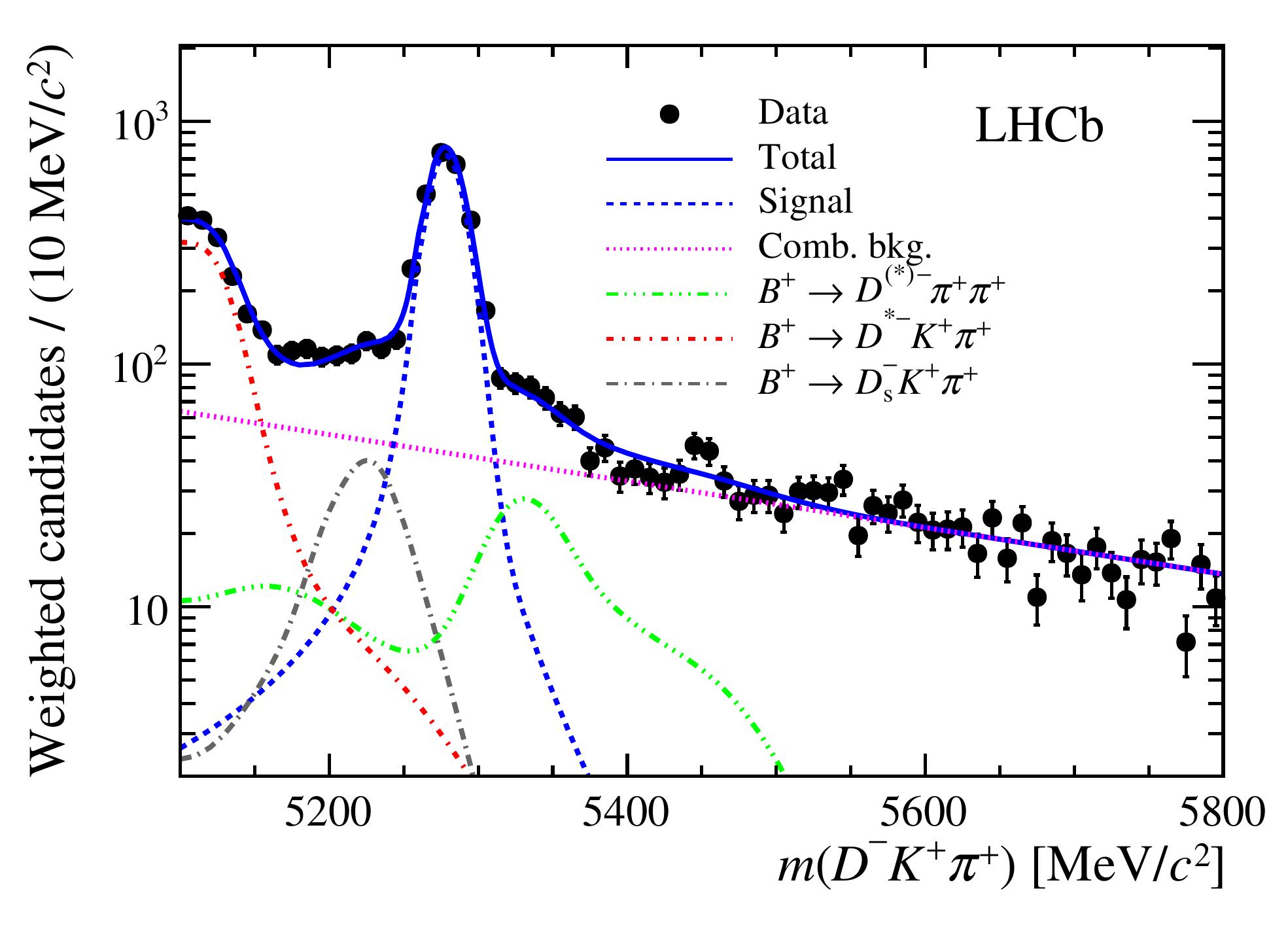}
\includegraphics[width=0.49\textwidth]{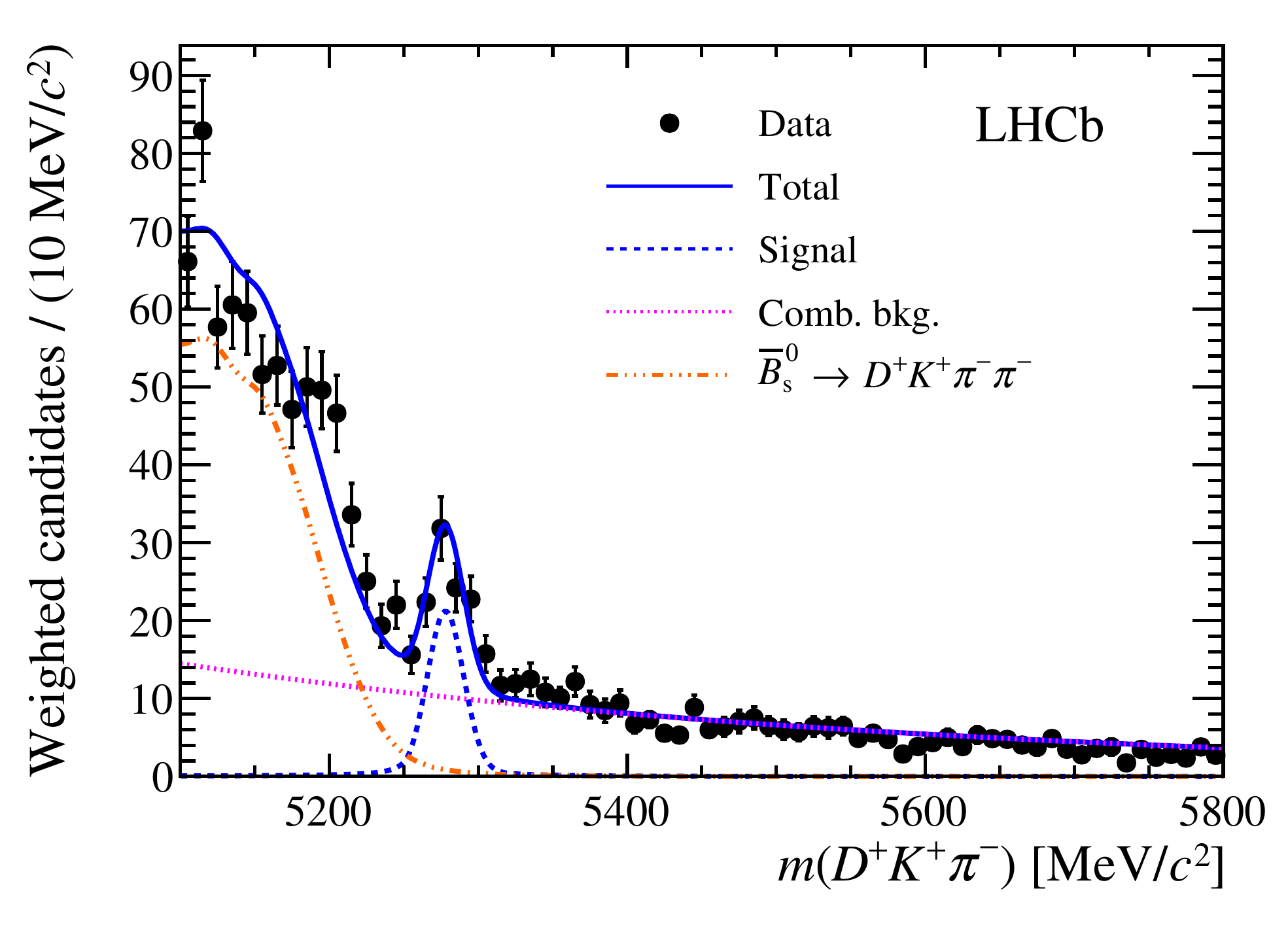}
\caption{\small
  Weighted invariant mass distribution of candidates in the (left) $\BtoRSdkpi$ and (right) $\BtoWSdkpi$ data samples. Data points and fit functions are weighted as described in the text.
 The components are as detailed in the legend.
}
\label{fig:fits}
\end{figure}

A total of 25 parameters are determined from the fit to the $\RSdkpi$ sample.
These include yields of $\BtoRSdkpi$ decays, the combinatorial background, the partially reconstructed background, and the $\Btodanddstarpipi$ and $\Btodskpi$ misidentified backgrounds.
For the signal category, and for combinatorial and partially reconstructed backgrounds, the fractional yields $f_i$ of each component in NN2 bins 1--5 are free parameters, with the fraction in bin 6 determined as $f_6 = 1-\sum_{i=1}^5 f_{i}$.
In addition, the exponential slope parameter of the combinatorial background and parameters of the signal invariant mass shape (the peak position, the width of the core CB function, the relative normalisation and ratio of the CB widths) are allowed to vary.
Figure~\ref{fig:fits} shows the combined \Bp candidate mass distribution of all NN2 bins weighted by $S/(S+B)$, where $S$ and $B$ are the fitted signal and background yields within $\pm 2.5\,\sigma_{\rm CB}$ of the signal peak position and $\sigma_{\rm CB}$ is the width of the core CB function.
% The reduced \chisq value obtained from the projections in each of the NN2 bins is 0.74.
The fit results are summarised in Table~\ref{tab:yields}.

The model for the fit to $\BtoWSdkpi$ candidates is similar to that for the $\BtoRSdkpi$ case.
The functional forms for the mass shapes for signal and combinatorial background are identical. The signal peak position, the width of the core CB function, and the fractional yields in each NN2 bin are fixed to the values obtained from the $\BtoRSdkpi$ fit.
A component is included for partially reconstructed background, which is likely to be dominated by $\Bstodkpipi$ decays; although this channel is unobserved, it is expected to be a sizable source of background based on studies of similar decay modes~\cite{LHCb-PAPER-2012-033,LHCb-PAPER-2014-035,LHCb-PAPER-2014-036}.
As the resonant structure of this mode is unknown, its mass shape is modelled using a combination of simulated samples generated with various $\Dp\pim$, $\Dp\Kp\pim$ and $\Kp\pim$ resonances and nonresonant amplitudes. 
The unknown structure of this background could cause some disagreement between data and the fit result at low $m(\Dp\Kp\pim)$.
The fractional yields in each NN2 bin are fixed to be the same as those for partially reconstructed backgrounds in the $\BtoRSdkpi$ fit.
Potential partially reconstructed background from \Bd and \Bu decays with a missing pion hardly enter the fit region; any residual contributions are absorbed in the \Bstodkpipi mass shape. 

There remain 11 parameters that are varied in the fit to the $\Dp\Kp\pim$ sample: the yields for $\BtoWSdkpi$ decays, combinatorial and partially reconstructed backgrounds; the fractional yields of the combinatorial background in each NN2 bin; the exponential slope parameter of the combinatorial background and the relative normalisation and ratio of widths of the two CB functions.
The results of this fit are summarised in Table~\ref{tab:yields} and shown in Fig.~\ref{fig:fits}.
% The projections of this fit in each of the NN2 bins have a reduced \chisq of 0.81.
The statistical significance of the $\WSdkpi$ peak, obtained from the square root of twice the change in negative log likelihood from the value obtained in a fit with zero signal yield, is $11\,\sigma$.

\begin{table}[!tb]
\centering
\caption{\small
  Yields and statistical uncertainties obtained from fits to the $\RSdkpi$ and $\WSdkpi$ data samples.}
\label{tab:yields}
\begin{tabular}{lr@{$\;\pm\;$}lr@{$\;\pm\;$}l}
\hline \\ [-2.5ex]
 & \multicolumn{2}{c}{$\RSdkpi$} & \multicolumn{2}{c}{$\WSdkpi$} \\ [0.2ex]
\hline \\ [-2.5ex]
$\BtoWSdkpi$ 		& \multicolumn{2}{c}{---} & $\phantom{11}164$ & $21$ \\ [0.2ex]
$\BtoRSdkpi$ 		& $3101$ & $66$ 	& \multicolumn{2}{c}{---} \\
Combinatorial background	& $3710$ & $110$ 	& $5945$ & $89$ \\
Partially reconstructed background& $1676$ & $57$ 	& $1425$ & $54$ \\
$\Btodanddstarpipi$ 	& $\phantom{11}548$ 	& $67$ & \multicolumn{2}{c}{---} \\
$\Btodskpi$ 		& $\phantom{11}342$ 	& $42$ & \multicolumn{2}{c}{---} \\
\hline
\end{tabular}
\end{table}

Systematic uncertainties on the ratio ${\BR\left(\decay{\Bp}{\WSdkpi}\right)}/{\BR\left(\decay{\Bp}{\RSdkpi}\right)}$ arise due to approximations made in the signal and background shapes used in the fit and uncertainties in the relative efficiencies.
The largest uncertainties are associated with the particle identification and hardware trigger efficiencies ($5.9\,\%$), the modelling of the combinatorial background in \B candidate mass and NN2 bins ($4.4\,\%$) and the NN2 distributions of signal and partially reconstructed background ($4.2\,\%$).
Other sources, including the modelling of the $\Bs\to\Dp\Km\pip\pim$ background and potential biases that are either intrinsic to the fit procedure or related to the treatment of multiple candidates, contribute systematic uncertainties of $2.0\,\%$ or less.

With all sources combined in quadrature, the total systematic uncertainty on the ratio of branching fractions is found to be $9.0\,\%$.
The likelihood function is convolved with a Gaussian of width corresponding to the size of the systematic uncertainties that affect the signal yield, and the total significance of the signal is found to be $8\,\sigma$.

The relative branching fraction of $\BtoWSdkpi$ and $\BtoRSdkpi$ decays is determined from
\begin{equation}
  \label{eqn:bfratio}
  \frac{\BR\left( \BtoWSdkpi \right)}{\BR\left( \BtoRSdkpi \right)} = \frac{N^{\rm corr}\left( \BtoWSdkpi \right)}{N^{\rm corr}\left( \BtoRSdkpi \right)}\, ,
\end{equation}
where the efficiency-corrected yield is $N^{\rm corr} = \sum_i W_i / \epsilon_i$.
Here the index $i$ runs over all candidates in the fit range,
$W_i$ is the signal weight for candidate $i$, determined using the \sPlot\ procedure~\cite{Pivk:2004ty}, from the fits shown in Fig.~\ref{fig:fits},
and $\epsilon_i$ is the efficiency for candidate $i$ as a function of its Dalitz plot position.

The average efficiencies are defined as $\bar{\epsilon} = N/N^{\rm corr} = \sum_i W_i/N^{\rm corr}$ and are found to be $\bar{\epsilon}\left( \BtoWSdkpi \right) = (0.057 \pm 0.014)\,\%$ and $\bar{\epsilon}\left( \BtoRSdkpi \right) = (0.079 \pm 0.003)\,\%$.
These values include contributions from the LHCb detector acceptance, selection and trigger. 
The trigger efficiency and most selection efficiencies are calculated from simulated samples with data-driven corrections applied, while the particle identification efficiency is measured from a data control sample~\cite{LHCb-DP-2012-003}.
The difference between the efficiencies is mainly caused by the different Dalitz plot distributions of the data in each channel.

From Eq.~(\ref{eqn:bfratio}), the ratio of branching fractions is determined to be
\begin{equation*}
\frac{\BR\left( \BtoWSdkpi \right)}{\BR\left( \BtoRSdkpi \right)} = 0.073 \pm 0.012 \stat \pm 0.007 \syst \, .
\end{equation*}
Taking $\BR\left(\BtoRSdkpi\right) = (7.31 \pm 0.19 \pm 0.22 \pm 0.39)\times 10^{-5}$~\cite{LHCb-PAPER-2015-007} gives
\begin{equation*}
\BR\left( \BtoWSdkpi \right) = ( 5.31 \pm 0.90 \pm 0.48 \pm 0.35)\times 10^{-6} \, ,
\end{equation*}
where the third uncertainty is from $\BR\left( \BtoRSdkpi \right)$, which arises mainly from the precision with which $\BR\left(\Bp\to\Dm\pip\pip\right)$~\cite{PDG2014} is known.

\begin{figure}[!b]
\centering
  \includegraphics[scale=0.45]{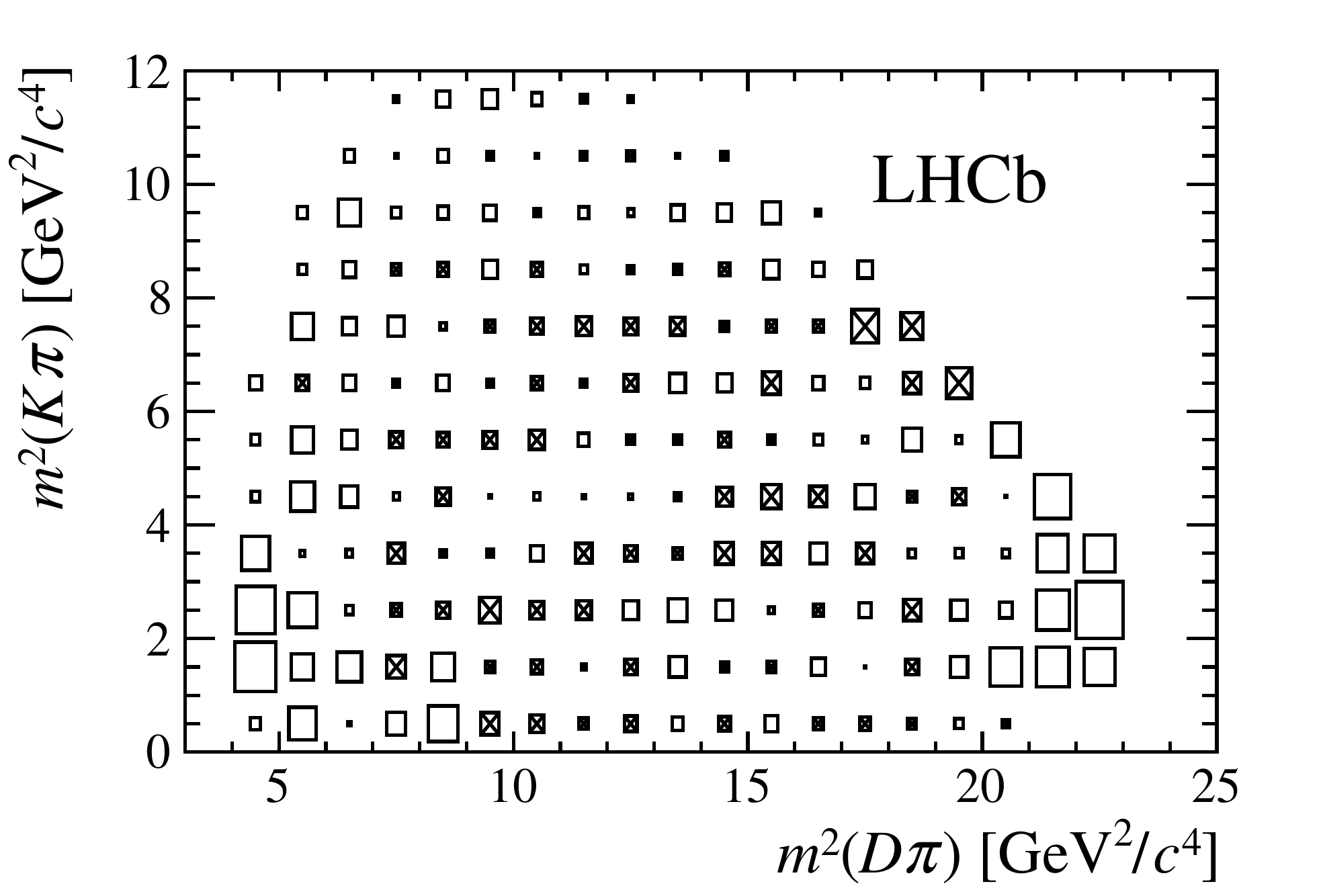}
\caption{\small 
  Background-subtracted Dalitz plot distribution of $\BtoWSdkpi$ candidates in the region $5260 < m(\WSdkpi) < 5310\mevcc$ from all NN2 bins.
  Areas of boxes are proportional to signal yields. Negative yields are indicated by crosses.
} 
\label{fig:WSdkpi-DP}
\end{figure}

The Dalitz plot distribution of $\BtoWSdkpi$ candidates in the region $5260 <\nobreak m(\WSdkpi) <\nobreak 5310\mevcc$ is shown in Fig.~\ref{fig:WSdkpi-DP}.
Combinatorial background has been subtracted using the distribution of candidates in a sideband ($5400 < m(\WSdkpi) < 5800\mevcc$), while the signal region has been chosen to minimise the $\Bstodkpipi$ background contribution.
The Dalitz plot variables are calculated with a constraint imposed on the \B mass; the combinatorial background distribution is not significantly distorted by this procedure.
Some excesses are seen at low $m(D\pi)$ and low $m(K\pi)$, but these do not appear to be from narrow structures; rather, there seems to be a broad S-wave $D\pi$ contribution.
The apparent structure at high $m(D\pi)$ may arise from imperfect background subtraction.

Although the $\BtoWSdkpi$ yield is not sufficient for a Dalitz plot analysis, it is possible to gain information about the contributions from narrow resonances.
Two-body mass requirements can reduce the contributions from other intermediate states, but not to a negligible level.
Therefore it is necessary to use the angular decay distributions to isolate particular resonances.
The $\BtoRSdkpi$ normalisation mode is again used to reduce potential sources of systematic bias.

Contributions from different partial waves can be determined by weighting the data according to the value of the Legendre polynomial of order $L$, $P_L$, evaluated as a function of the cosine of the helicity angle of the $\Kp\pim$ or $\Dp\pim$ system.
The helicity angle is defined as the angle between the momentum vectors of the pion and the \Bp candidate in the $\Kp\pim$ or $\Dp\pim$ rest frame. 
Event-by-event efficiency corrections, determined as a function of Dalitz plot position, are also applied.
The helicity angles and two-body invariant masses are calculated with a constraint on the \B mass imposed on the decay chain.  
If only resonances up to spin $J_{\rm max}$ are present in a certain mass region, the $P_{2J_{\rm max}}$ moment will isolate the highest spin state.
Thus, in the limit that only $\Dp\pim$ resonances contribute, weighting by $P_4$ will isolate the $D_2^*(2460)^0$ component, as shown in Ref.~\cite{LHCb-PAPER-2015-007}, where a more detailed description of contributions to each moment can be found.
Similarly, at low $m(\Kp\pim)$, weighting by $P_2$ can be used to determine the contribution from the $K^*(892)^0$ resonance.
Higher moments may be present, due to tails of higher spin resonances or reflections from resonances in the other two-particle combination; these will also cause an excess of events in regions away from the resonance peak and therefore can be accounted for by sideband subtraction.

Candidates are selected within regions corresponding to approximately $\pm 2\Gamma$, where $\Gamma$ is the natural width~\cite{PDG2014}, around the peaks of the $\DorDbar_2^*(2460)^0$ resonance in $m(\Dpm\pimp)$ and of the $K^*(892)^0$ resonance in $m(\Kp\pim)$.
The data are efficiency-corrected and weighted according to the corresponding Legendre polynomial functions.
Yields, denoted $\widetilde{N}$, are then obtained from binned minimum $\chisq$ fits to the \Bp candidate mass distribution.
A variable-width binning scheme is used with bin widths chosen to avoid empty bins.
The same procedure is applied for candidates in low and high sideband regions, between about $3\Gamma$ and $5\Gamma$ from the peak.
For the normalisation of the search for $\Dp K^*(892)^0$ decays, the full
efficiency-corrected $\Dm\Kp\pip$ sample is used without weighting by angular
moment.
The results are used to measure the ratios of branching fractions
\begin{eqnarray}
  && \hspace{-10mm} 
  \frac{\BR\left(\decay{\Bp}{D^*_2(2460)^0 K^+}\right)}{\BR\left(\decay{\Bp}{\Dbar{}^*_2(2460)^0 K^+}\right)} \equiv \left(r_B(D^*_2(2460)K^+)\right)^2 =
  \frac{\widetilde{N}^{\rm corr}\left(\decay{\Bp}{D^*_2(2460)^0 K^+}\right)}{\widetilde{N}^{\rm corr}\left(\decay{\Bp}{\Dbar{}^*_2(2460)^0 K^+}\right)} \, , \label{eq:pfour} \\
  && \hspace{-10mm} 
  \frac{\BR\left(\decay{\Bp}{\Dp K^*(892)^0 \to \Dp\Kp\pim}\right)}{\BR\left(\decay{\Bp}{\RSdkpi}\right)} =
  \frac{\widetilde{N}^{\rm corr}\left(\decay{\Bp}{\Dp K^*(892)^0}\right) \cdot \left( \frac{5}{2} \right)}{\widetilde{N}^{\rm corr}\left(\decay{\Bp}{\RSdkpi}\right) \epsilon(K^*(892)^0)} \, \, , \label{eq:ptwo} %\BR\left(K^*(892)^0\right)} 
\end{eqnarray}
where $\widetilde{N}^{\rm corr}$ are the yields obtained from the fit after accounting for subtraction of higher moments as estimated from the sideband regions.
In Eq.~(\ref{eq:ptwo}) the correction of $\frac{5}{2}$ arises from the normalisation of the Legendre polynomial functions and the factor of $\epsilon(K^*(892)^0) = 0.857 \pm 0.006$ is due to the efficiency of the $K^*(892)^0$ signal region ($801.0 < m(\Kp\pim) < 990.6\mevcc$) requirement.
All efficiency, $D^*_2$ branching fraction and normalisation effects cancel in Eq.~(\ref{eq:pfour}).

The fit models used are based on those described above, but with some important simplifications.
The angular weighting by $P_2$ or $P_4$ significantly reduces the combinatorial background, and therefore candidates in all NN2 bins are combined; moreover a linear shape is used instead of an exponential function in order to allow for the possibility that the weighted background can fluctuate to negative values.
The $\Btodanddstarpipi$ and $\Bstodkpipi$ background shapes are given by non-parametric functions obtained from simulated samples with angular moment weighting applied.
No component is included for misidentified $\Btodskpi$ decays, as it is found to be removed by the weighting procedure. 

The analysis method is validated using the $\BtoRSdkpi$ channel and simulated pseudo\-experiments.
The fit fraction for $\decay{\Bp}{\Dbar{}^*_2(2460)^0 K^+}$ decays obtained from a full Dalitz plot analysis in Ref.~\cite{LHCb-PAPER-2015-007} is reproduced within the expected range.
The procedure is tested by searching for a fake $\Kstar$ resonance in $m(\Kp\pip)$, and the yield is found to be consistent with zero.

\begin{figure}[!tb]
  \centering
  \includegraphics[scale=0.35]{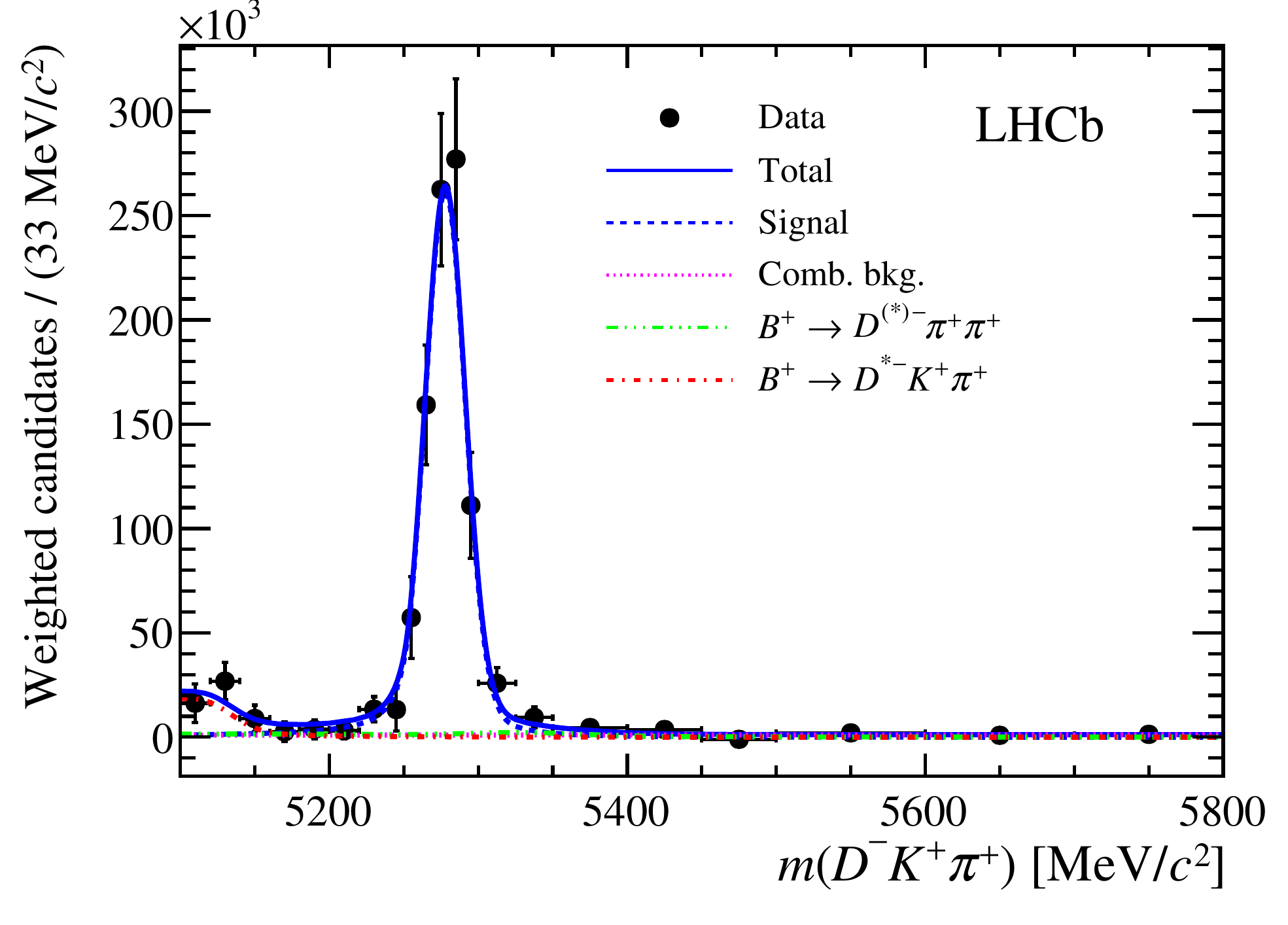}
  \includegraphics[scale=0.35]{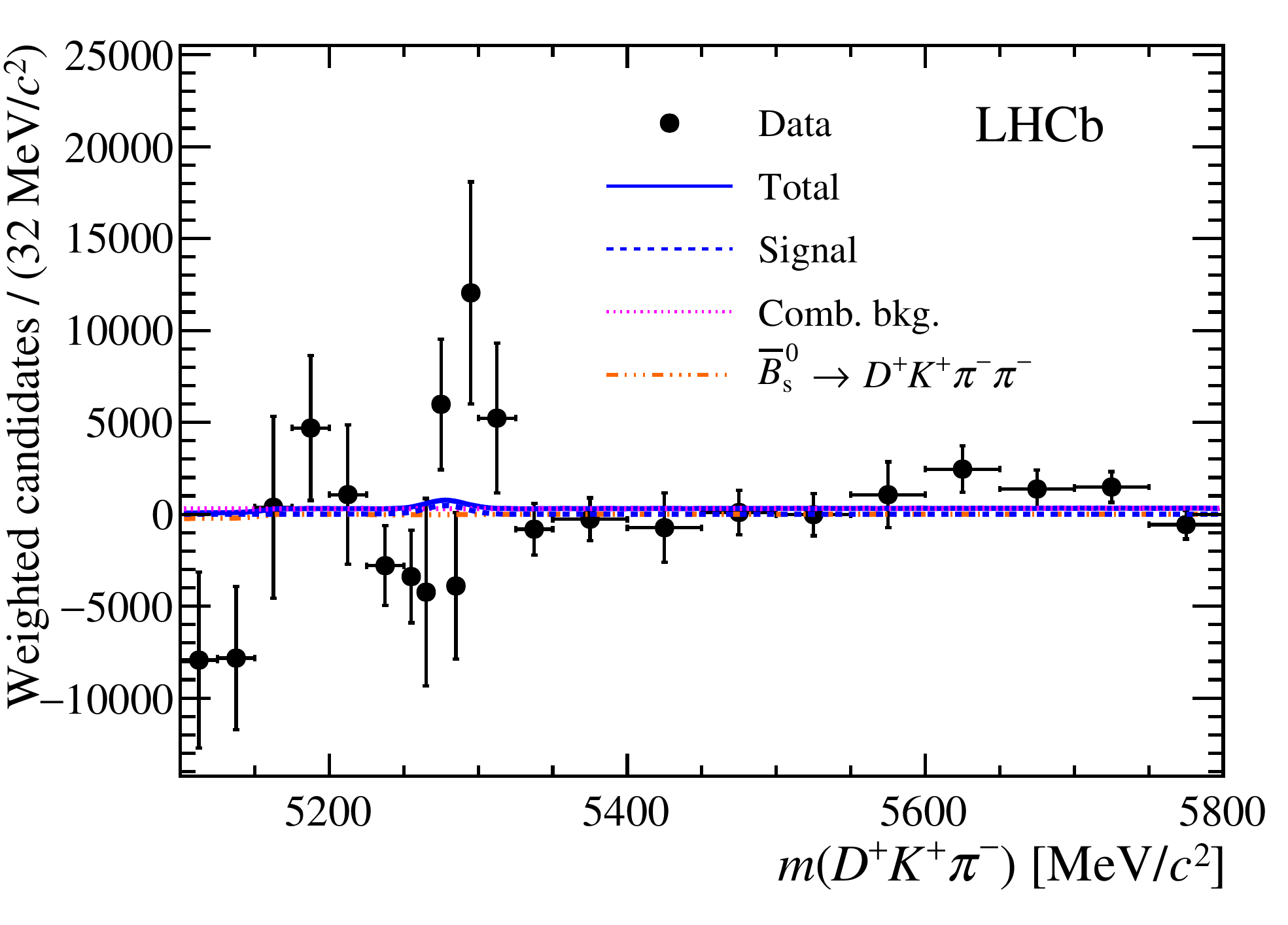}
  \includegraphics[scale=0.35]{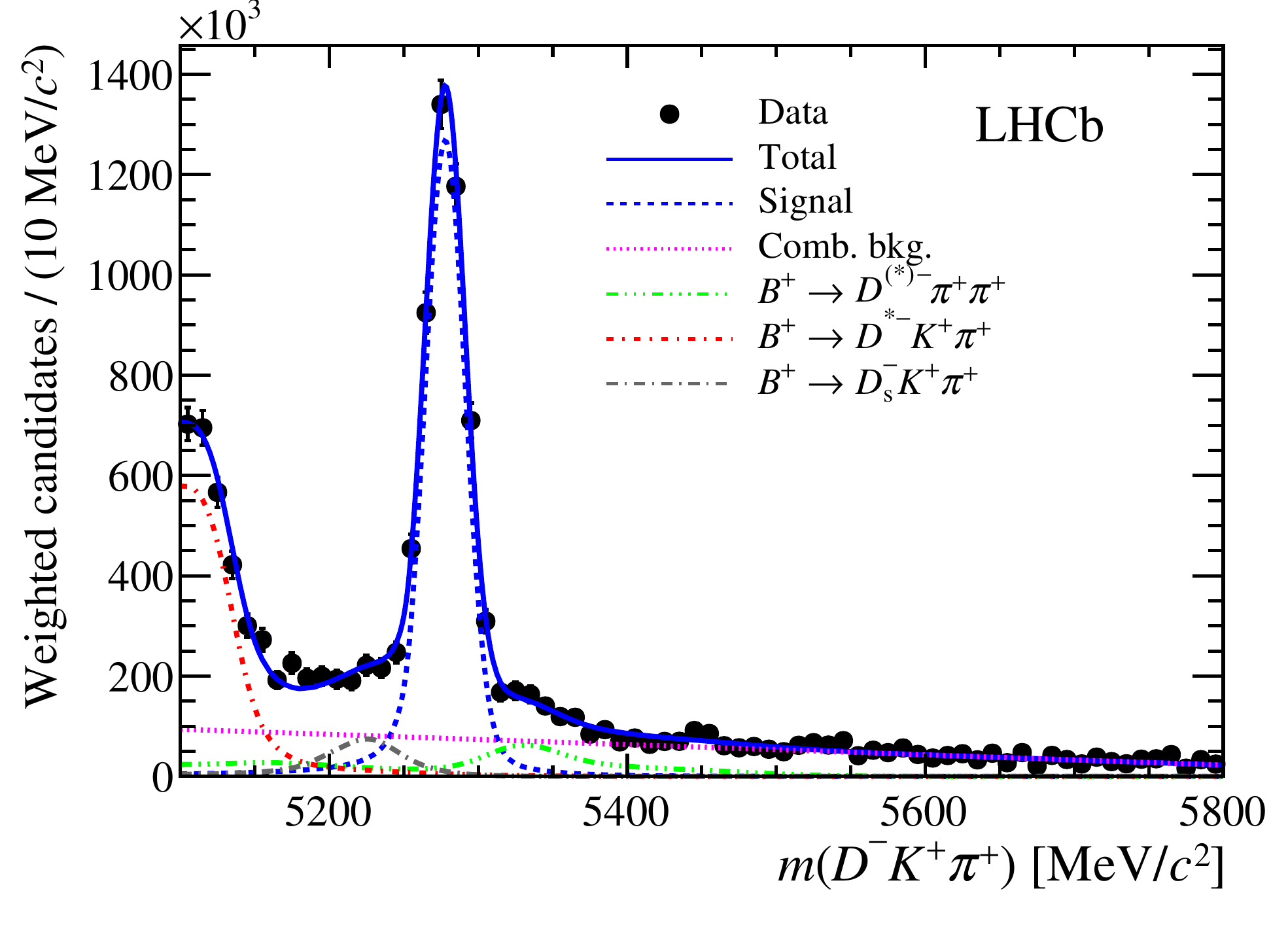}
  \includegraphics[scale=0.35]{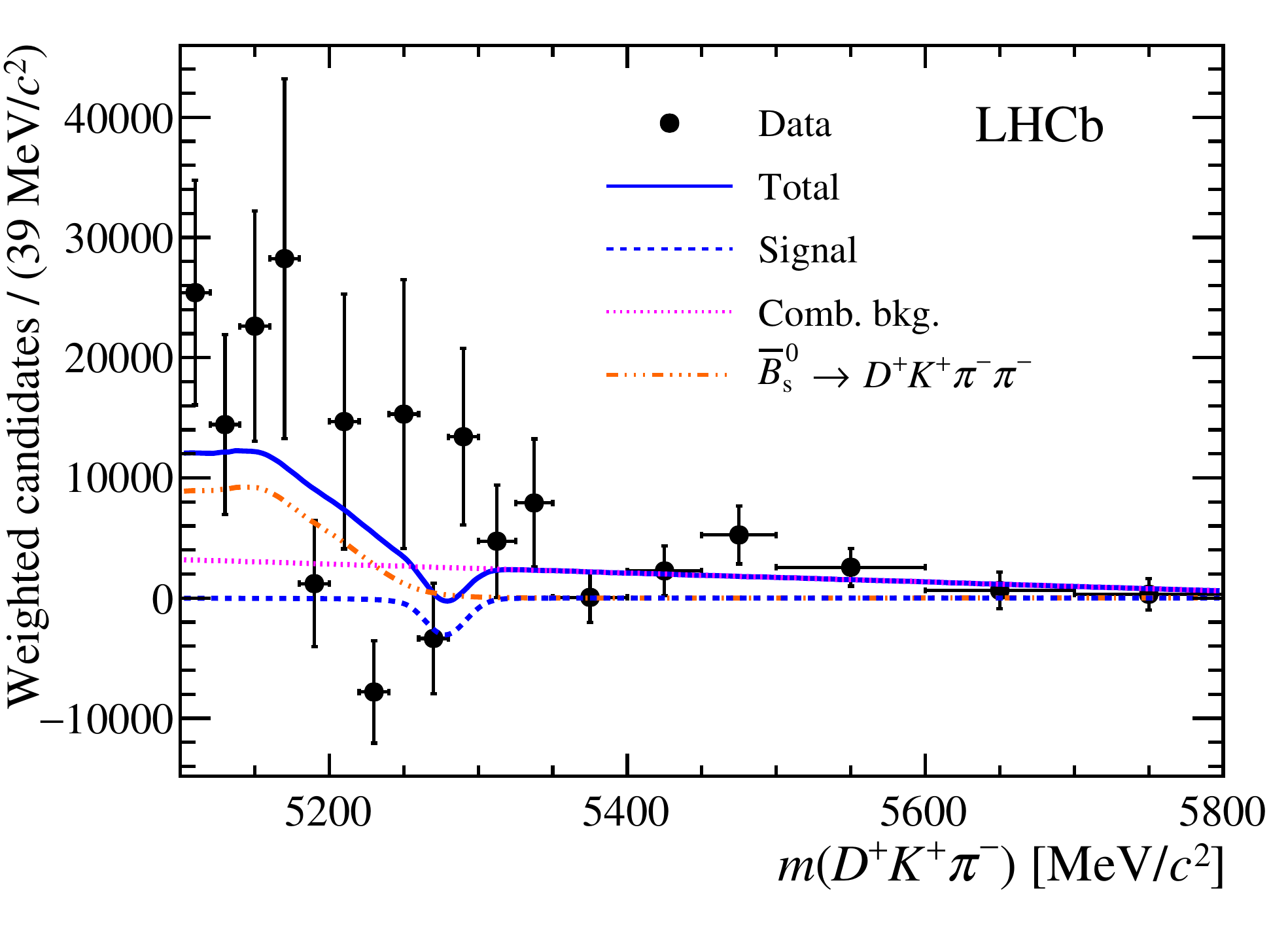}
  \caption{\small 
    Results of binned minimum $\chisq$ fits to the \Bp candidate invariant mass distributions of (left) \BtoRSdkpi and (right) \BtoWSdkpi candidates with efficiency corrections and angular weights applied. 
    Candidates in the ranges (top) $2217.6 < m(D^\pm\pi^\mp) < 2315.6\mevcc$ and
    (bottom right) $801.0 < m(\Kp\pim) < 990.6\mevcc$ are shown, while in the bottom left plot the whole $m(\Kp\pip)$ range is included.
    The components are as described in the legend.
}
\label{fig:twobody}
\end{figure}

The results of the fits to $P_4$-weighted and efficiency-corrected $\BtoWSdkpi$ and $\BtoRSdkpi$ data samples in the $\DorDbar_2^*(2460)^0$ resonance region are shown in Fig.~\ref{fig:twobody}.
The procedure isolates the $\decay{\Bp}{\Dbar{}^*_2(2460)^0 K^+}$ decay, as expected, but no evidence is seen for the suppressed $\decay{\Bp}{D^*_2(2460)^0 K^+}$ channel.
The corresponding fits for the $\BtoKstarD$ search are also shown in Fig.~\ref{fig:twobody}; there is no evidence for this decay.
The yields are given in Table~\ref{tab:twobody-results}.

\begin{table}[!tb]
  \centering
  \caption{
    \small Results of the binned minimum $\chisq$ fits to efficiency-corrected $\Bp$ candidate invariant mass distributions in each resonance region and with weighting according to angular distributions as described in the text.
  }
  \label{tab:twobody-results}
  \resizebox{\textwidth}{!}{
  \begin{tabular}{cr@{$\,\pm\,$}l@{\hspace{-5mm}}r@{$\,\pm\,$}lr@{$\,\pm\,$}l@{\hspace{-5mm}}r@{$\,\pm\,$}l}
    \hline \\ [-2.3ex]
    & \multicolumn{2}{c}{Lower sideband} & \multicolumn{2}{c}{Signal region} &
    \multicolumn{2}{c}{Upper sideband}  & \multicolumn{2}{c}{$\widetilde{N}^{\rm corr}$} \\
    \hline \\ [-2.3ex]
    $\widetilde{N}(\Bp \to D_2^*(2460)^0\Kp)$       & $-200$    & $\phani2\,500$ &      $500$    & $\phanii3\,000$ &  $200$ & $2\,200$       &      $500$ & $\phani4\,500$ \\ 
    $\widetilde{N}(\Bp \to \Dbar{}_2^*(2460)^0\Kp)$ & $28\,000$ &      $14\,000$ & $293\,000$    & $\phani24\,000$ & $-600$ & $4\,200$       & $266\,000$ & $     28\,000$ \\
    \hline \\ [-2.3ex]
    $\widetilde{N}(\Bp \to \Dp K^*(892)^0)$         & $1\,700$  & $\phani1\,900$ & $-3\,000$     & $\phanii5\,000$ & $9\,500$ & $4\,000$     & $-14\,000$ & $\phani7\,000$ \\
    $\widetilde{N}(\BtoRSdkpi)$                     & \multicolumn{2}{c}{---}    & $4\,670\,000$ &      $110\,000$ & \multicolumn{2}{c}{---} & \multicolumn{2}{c}{---} \\
    \hline
  \end{tabular}
}
\end{table}

Systematic uncertainties arise due to the fit models and background subtraction used to determine $\widetilde{N}^{\rm corr}$ in Eqs.~(\ref{eq:pfour}) and~(\ref{eq:ptwo}).
The uncertainties are evaluated from the effects on the yields of the following variations:
the combinatorial background shape is changed from linear to flat;
the $\Bstodkpipi$ background component is removed;
all other fit components are varied in the same way as described previously.
The limited precision of the knowledge of the efficiencies as functions of Dalitz plot position also causes a small uncertainty.
An uncertainty is assigned due to the effect of changing the sideband regions from the default of $3\Gamma \leftrightarrow 5\Gamma$ to $4\Gamma \leftrightarrow 6\Gamma$.
The uncertainty in $\epsilon(K^*(892)^0)$ of Eq.~(\ref{eq:ptwo}) is also accounted for.
The total systematic uncertainty is obtained by combining all sources in quadrature.

The ratio of branching fractions is thus measured to be 
\begin{equation*}
\frac{\BR\left(\decay{\Bp}{D^*_2(2460)^0 K^+}\right)}{\BR\left(\decay{\Bp}{\Dbar{}^*_2(2460)^0 K^+}\right)} = 0.002 \pm 0.015 \stat \pm 0.005 \syst \, ,
\end{equation*}
which in turn gives
\begin{equation*}
r_B(D^*_2(2460)K^+) = 0.04 \pm 0.18 \stat \pm 0.06 \syst \, .
\end{equation*}
Assuming Gaussian uncertainties, upper limits at $90\,(95)\,\%$ confidence level (CL) are obtained by integrating the likelihood in the region of positive branching fraction,
\begin{equation*}
\left(r_B(D^*_2(2460)K^+)\right)^2 < 0.027~(0.033) 
\quad {\rm and} \quad
r_B(D^*_2(2460)\Kp) < 0.30~(0.36) \, .
\end{equation*}

The result for $\left(r_B(D^*_2(2460)K^+)\right)^2$ and the product branching fraction $\BR\left(\decay{\Bp}{\Dbar{}^*_2(2460)^0 K^+}\right) \times \BR\left(\decay{\Dbar{}^*_2(2460)^0}{\Dm\pip}\right) = (23.2 \pm 1.1 \pm 0.6 \pm 1.0 \pm 1.6) \times 10^{-6}$~\cite{LHCb-PAPER-2015-007} 
give
\begin{eqnarray*}
  \BR\left(\decay{\Bp}{D^*_2(2460)^0 K^+}\right) \times \BR\left(\decay{D^*_2(2460)^0}{\Dp\pim}\right) \hspace{-1mm} & = & \hspace{-1mm} (0.4 \pm 3.5 \pm 1.1 \pm 0.1) \times 10^{-7} \, , \\
  & < & \hspace{-1mm} 6.3~(7.5) \times 10^{-7} \ {\rm at} \ 90\,(95)\,\% \ {\rm CL.}
\end{eqnarray*}
These are the first experimental results on this decay mode.
Similarly for $\decay{\Bp}{\Dp K^*(892)^0 \to \Dp\Kp\pim}$, 
\begin{eqnarray*}
  \frac{\BR\left(\decay{\Bp}{\Dp K^*(892)^0 \to \Dp\Kp\pim}\right) }{\BR\left(\decay{\Bp}{\RSdkpi}\right)} & = & -0.0079 \pm 0.0039 \stat \pm 0.0028 \syst \, , \\ 
  & < & 0.0044~(0.0055) \ {\rm at} \ 90\,(95)\,\% \ {\rm CL.} 
\end{eqnarray*}
The measured value ${\cal B}\left(\Bp \to \Dm \Kp\pip\right) = (7.31 \pm 0.19 \pm 0.22 \pm 0.39)\times 10^{-5}$~\cite{LHCb-PAPER-2015-007} and the isospin relation $\BR\left(\decay{K^*(892)^0}{\Kp\pim} \right) = \frac{2}{3}$ give
\begin{eqnarray*}
  \BR\left(\decay{\Bp}{\Dp K^*(892)^0}\right) & = & (-8.7 \pm 4.3 \pm 3.1 \pm 0.4) \times 10^{-7} \, , \\
  & < &  4.9~(6.1) \times 10^{-7} \ {\rm at} \ 90\,(95)\,\% \ {\rm CL,}
\end{eqnarray*}
where the third uncertainty is due to the normalisation channel branching fraction.
This result supersedes the previous limit, which was obtained with a subset of the data~\cite{LHCb-PAPER-2012-025}.

In summary, the rare $\BtoWSdkpi$ decay has been observed for the first time with $8\,\sigma$ significance, based on a data sample of $3.0\invfb$ of $pp$ collision data collected with the LHCb detector.
The Dalitz plot appears to be dominated by broad structures.
Searches for $\Bp \to D_2^*(2460)^0\Kp$ and $\Bp \to \Dp K^*(892)^0$ decays have been carried out by weighting the data according to the decay angle distributions, but no significant signals are seen.
These results indicate that further studies, with larger data samples, of the Dalitz plot distribution of this mode will be of interest to understand the potential for a measurement of $\gamma$ from $\Bp\to \D\Kp\piz$ decays.

\section*{Acknowledgements}

\noindent We express our gratitude to our colleagues in the CERN
accelerator departments for the excellent performance of the LHC. We
thank the technical and administrative staff at the LHCb
institutes. We acknowledge support from CERN and from the national
agencies: CAPES, CNPq, FAPERJ and FINEP (Brazil); NSFC (China);
CNRS/IN2P3 (France); BMBF, DFG and MPG (Germany); INFN (Italy); 
FOM and NWO (The Netherlands); MNiSW and NCN (Poland); MEN/IFA (Romania); 
MinES and FANO (Russia); MinECo (Spain); SNSF and SER (Switzerland); 
NASU (Ukraine); STFC (United Kingdom); NSF (USA).
We acknowledge the computing resources that are provided by CERN, IN2P3 (France), KIT and DESY (Germany), INFN (Italy), SURF (The Netherlands), PIC (Spain), GridPP (United Kingdom), RRCKI (Russia), CSCS (Switzerland), IFIN-HH (Romania), CBPF (Brazil), PL-GRID (Poland) and OSC (USA). We are indebted to the communities behind the multiple open 
source software packages on which we depend. We are also thankful for the 
computing resources and the access to software R\&D tools provided by Yandex LLC (Russia).
Individual groups or members have received support from AvH Foundation (Germany),
EPLANET, Marie Sk\l{}odowska-Curie Actions and ERC (European Union), 
Conseil G\'{e}n\'{e}ral de Haute-Savoie, Labex ENIGMASS and OCEVU, 
R\'{e}gion Auvergne (France), RFBR (Russia), GVA, XuntaGal and GENCAT (Spain), The Royal Society 
and Royal Commission for the Exhibition of 1851 (United Kingdom).

\ifx\mcitethebibliography\mciteundefinedmacro
\PackageError{LHCb.bst}{mciteplus.sty has not been loaded}
{This bibstyle requires the use of the mciteplus package.}\fi
\providecommand{\href}[2]{#2}

\addcontentsline{toc}{section}{References}
\setboolean{inbibliography}{true}
%\bibliographystyle{LHCb}
%\bibliography{main,references,LHCb-PAPER,LHCb-CONF,LHCb-DP,LHCb-TDR}

\begin{mcitethebibliography}{10}
\mciteSetBstSublistMode{n}
\mciteSetBstMaxWidthForm{subitem}{\alph{mcitesubitemcount})}
\mciteSetBstSublistLabelBeginEnd{\mcitemaxwidthsubitemform\space}
{\relax}{\relax}

\bibitem{Cabibbo:1963yz}
N.~Cabibbo, \ifthenelse{\boolean{articletitles}}{\emph{{Unitary symmetry and
  leptonic decays}},
  }{}\href{http://dx.doi.org/10.1103/PhysRevLett.10.531}{Phys.\ Rev.\ Lett.\
  \textbf{10} (1963) 531}\relax
\mciteBstWouldAddEndPuncttrue
\mciteSetBstMidEndSepPunct{\mcitedefaultmidpunct}
{\mcitedefaultendpunct}{\mcitedefaultseppunct}\relax
\EndOfBibitem
\bibitem{Kobayashi:1973fv}
M.~Kobayashi and T.~Maskawa, \ifthenelse{\boolean{articletitles}}{\emph{{\CP
  violation in the renormalizable theory of weak interaction}},
  }{}\href{http://dx.doi.org/10.1143/PTP.49.652}{Prog.\ Theor.\ Phys.\
  \textbf{49} (1973) 652}\relax
\mciteBstWouldAddEndPuncttrue
\mciteSetBstMidEndSepPunct{\mcitedefaultmidpunct}
{\mcitedefaultendpunct}{\mcitedefaultseppunct}\relax
\EndOfBibitem
\bibitem{Bona:2005vz}
UTfit collaboration, M.~Bona {\em et~al.},
  \ifthenelse{\boolean{articletitles}}{\emph{{The 2004 UTfit collaboration
  report on the status of the unitarity triangle in the standard model}},
  }{}\href{http://dx.doi.org/10.1088/1126-6708/2005/07/028}{JHEP \textbf{07}
  (2005) 028}, \href{http://arxiv.org/abs/hep-ph/0501199}{{\normalfont\ttfamily
  arXiv:hep-ph/0501199}}\relax
\mciteBstWouldAddEndPuncttrue
\mciteSetBstMidEndSepPunct{\mcitedefaultmidpunct}
{\mcitedefaultendpunct}{\mcitedefaultseppunct}\relax
\EndOfBibitem
\bibitem{Charles:2004jd}
CKMfitter group, J.~Charles {\em et~al.},
  \ifthenelse{\boolean{articletitles}}{\emph{{CP violation and the CKM matrix:
  Assessing the impact of the asymmetric $B$ factories}},
  }{}\href{http://dx.doi.org/10.1140/epjc/s2005-02169-1}{Eur.\ Phys.\ J.\
  \textbf{C41} (2005) 1},
  \href{http://arxiv.org/abs/hep-ph/0406184}{{\normalfont\ttfamily
  arXiv:hep-ph/0406184}}\relax
\mciteBstWouldAddEndPuncttrue
\mciteSetBstMidEndSepPunct{\mcitedefaultmidpunct}
{\mcitedefaultendpunct}{\mcitedefaultseppunct}\relax
\EndOfBibitem
\bibitem{LHCb-CONF-2014-004}
{LHCb collaboration}, \ifthenelse{\boolean{articletitles}}{\emph{{Improved
  constraints on $\gamma$: CKM2014 update}}, }{}
  \href{http://cdsweb.cern.ch/search?p=LHCb-CONF-2014-004&f=reportnumber&action_search=Search&c=LHCb+Conference+Contributions}
  {LHCb-CONF-2014-004}\relax
\mciteBstWouldAddEndPuncttrue
\mciteSetBstMidEndSepPunct{\mcitedefaultmidpunct}
{\mcitedefaultendpunct}{\mcitedefaultseppunct}\relax
\EndOfBibitem
\bibitem{Gronau:1990ra}
M.~Gronau and D.~London, \ifthenelse{\boolean{articletitles}}{\emph{{How to
  determine all the angles of the unitarity triangle from $\Bz \to D\KS$ and
  $\Bs \to D\phi$}},
  }{}\href{http://dx.doi.org/10.1016/0370-2693(91)91756-L}{Phys.\ Lett.\
  \textbf{B253} (1991) 483}\relax
\mciteBstWouldAddEndPuncttrue
\mciteSetBstMidEndSepPunct{\mcitedefaultmidpunct}
{\mcitedefaultendpunct}{\mcitedefaultseppunct}\relax
\EndOfBibitem
\bibitem{Gronau:1991dp}
M.~Gronau and D.~Wyler, \ifthenelse{\boolean{articletitles}}{\emph{{On
  determining a weak phase from charged \B decay asymmetries}},
  }{}\href{http://dx.doi.org/10.1016/0370-2693(91)90034-N}{Phys.\ Lett.\
  \textbf{B265} (1991) 172}\relax
\mciteBstWouldAddEndPuncttrue
\mciteSetBstMidEndSepPunct{\mcitedefaultmidpunct}
{\mcitedefaultendpunct}{\mcitedefaultseppunct}\relax
\EndOfBibitem
\bibitem{Atwood:1996ci}
D.~Atwood, I.~Dunietz, and A.~Soni,
  \ifthenelse{\boolean{articletitles}}{\emph{{Enhanced \CP violation with $\B
  \to K \Dz (\Dzb)$ modes and extraction of the CKM angle $\gamma$}},
  }{}\href{http://dx.doi.org/10.1103/PhysRevLett.78.3257}{Phys.\ Rev.\ Lett.\
  \textbf{78} (1997) 3257},
  \href{http://arxiv.org/abs/hep-ph/9612433}{{\normalfont\ttfamily
  arXiv:hep-ph/9612433}}\relax
\mciteBstWouldAddEndPuncttrue
\mciteSetBstMidEndSepPunct{\mcitedefaultmidpunct}
{\mcitedefaultendpunct}{\mcitedefaultseppunct}\relax
\EndOfBibitem
\bibitem{Atwood:2000ck}
D.~Atwood, I.~Dunietz, and A.~Soni,
  \ifthenelse{\boolean{articletitles}}{\emph{{Improved methods for observing
  \CP violation in $\Bpm \to \Kpm D$ and measuring the CKM phase $\gamma$}},
  }{}\href{http://dx.doi.org/10.1103/PhysRevD.63.036005}{Phys.\ Rev.\
  \textbf{D63} (2001) 036005},
  \href{http://arxiv.org/abs/hep-ph/0008090}{{\normalfont\ttfamily
  arXiv:hep-ph/0008090}}\relax
\mciteBstWouldAddEndPuncttrue
\mciteSetBstMidEndSepPunct{\mcitedefaultmidpunct}
{\mcitedefaultendpunct}{\mcitedefaultseppunct}\relax
\EndOfBibitem
\bibitem{Sinha:2004ct}
N.~Sinha, \ifthenelse{\boolean{articletitles}}{\emph{{Determining $\gamma$
  using $B \to D^{**}K$}},
  }{}\href{http://dx.doi.org/10.1103/PhysRevD.70.097501}{Phys.\ Rev.\
  \textbf{D70} (2004) 097501},
  \href{http://arxiv.org/abs/hep-ph/0405061}{{\normalfont\ttfamily
  arXiv:hep-ph/0405061}}\relax
\mciteBstWouldAddEndPuncttrue
\mciteSetBstMidEndSepPunct{\mcitedefaultmidpunct}
{\mcitedefaultendpunct}{\mcitedefaultseppunct}\relax
\EndOfBibitem
\bibitem{LHCb-PAPER-2015-007}
LHCb collaboration, R.~Aaij {\em et~al.},
  \ifthenelse{\boolean{articletitles}}{\emph{{First observation and amplitude
  analysis of the $B^{-}\to D^{+}K^{-}\pi^{-}$ decay}},
  }{}\href{http://dx.doi.org/10.1103/PhysRevD.91.092002}{Phys.\ Rev.\
  \textbf{D91} (2015) 092002},
  \href{http://arxiv.org/abs/1503.02995}{{\normalfont\ttfamily
  arXiv:1503.02995}}\relax
\mciteBstWouldAddEndPuncttrue
\mciteSetBstMidEndSepPunct{\mcitedefaultmidpunct}
{\mcitedefaultendpunct}{\mcitedefaultseppunct}\relax
\EndOfBibitem
\bibitem{LHCb-PAPER-2012-025}
LHCb collaboration, R.~Aaij {\em et~al.},
  \ifthenelse{\boolean{articletitles}}{\emph{{First evidence for the
  annihilation decay mode $B^+ \to D_s^+\phi$}},
  }{}\href{http://dx.doi.org/10.1007/JHEP02(2013)043}{JHEP \textbf{02} (2013)
  043}, \href{http://arxiv.org/abs/1210.1089}{{\normalfont\ttfamily
  arXiv:1210.1089}}\relax
\mciteBstWouldAddEndPuncttrue
\mciteSetBstMidEndSepPunct{\mcitedefaultmidpunct}
{\mcitedefaultendpunct}{\mcitedefaultseppunct}\relax
\EndOfBibitem
\bibitem{Alves:2008zz}
LHCb collaboration, A.~A. Alves~Jr.\ {\em et~al.},
  \ifthenelse{\boolean{articletitles}}{\emph{{The \lhcb detector at the LHC}},
  }{}\href{http://dx.doi.org/10.1088/1748-0221/3/08/S08005}{JINST \textbf{3}
  (2008) S08005}\relax
\mciteBstWouldAddEndPuncttrue
\mciteSetBstMidEndSepPunct{\mcitedefaultmidpunct}
{\mcitedefaultendpunct}{\mcitedefaultseppunct}\relax
\EndOfBibitem
\bibitem{LHCb-DP-2014-002}
LHCb collaboration, R.~Aaij {\em et~al.},
  \ifthenelse{\boolean{articletitles}}{\emph{{LHCb detector performance}},
  }{}\href{http://dx.doi.org/10.1142/S0217751X15300227}{Int.\ J.\ Mod.\ Phys.\
  \textbf{A30} (2015) 1530022},
  \href{http://arxiv.org/abs/1412.6352}{{\normalfont\ttfamily
  arXiv:1412.6352}}\relax
\mciteBstWouldAddEndPuncttrue
\mciteSetBstMidEndSepPunct{\mcitedefaultmidpunct}
{\mcitedefaultendpunct}{\mcitedefaultseppunct}\relax
\EndOfBibitem
\bibitem{Sjostrand:2007gs}
T.~Sj\"{o}strand, S.~Mrenna, and P.~Skands,
  \ifthenelse{\boolean{articletitles}}{\emph{{A brief introduction to PYTHIA
  8.1}}, }{}\href{http://dx.doi.org/10.1016/j.cpc.2008.01.036}{Comput.\ Phys.\
  Commun.\  \textbf{178} (2008) 852},
  \href{http://arxiv.org/abs/0710.3820}{{\normalfont\ttfamily
  arXiv:0710.3820}}\relax
\mciteBstWouldAddEndPuncttrue
\mciteSetBstMidEndSepPunct{\mcitedefaultmidpunct}
{\mcitedefaultendpunct}{\mcitedefaultseppunct}\relax
\EndOfBibitem
\bibitem{Sjostrand:2006za}
T.~Sj\"{o}strand, S.~Mrenna, and P.~Skands,
  \ifthenelse{\boolean{articletitles}}{\emph{{PYTHIA 6.4 physics and manual}},
  }{}\href{http://dx.doi.org/10.1088/1126-6708/2006/05/026}{JHEP \textbf{05}
  (2006) 026}, \href{http://arxiv.org/abs/hep-ph/0603175}{{\normalfont\ttfamily
  arXiv:hep-ph/0603175}}\relax
\mciteBstWouldAddEndPuncttrue
\mciteSetBstMidEndSepPunct{\mcitedefaultmidpunct}
{\mcitedefaultendpunct}{\mcitedefaultseppunct}\relax
\EndOfBibitem
\bibitem{LHCb-PROC-2010-056}
I.~Belyaev {\em et~al.}, \ifthenelse{\boolean{articletitles}}{\emph{{Handling
  of the generation of primary events in Gauss, the LHCb simulation
  framework}}, }{}\href{http://dx.doi.org/10.1088/1742-6596/331/3/032047}{{J.\
  Phys.\ Conf.\ Ser.\ } \textbf{331} (2011) 032047}\relax
\mciteBstWouldAddEndPuncttrue
\mciteSetBstMidEndSepPunct{\mcitedefaultmidpunct}
{\mcitedefaultendpunct}{\mcitedefaultseppunct}\relax
\EndOfBibitem
\bibitem{Lange:2001uf}
D.~J. Lange, \ifthenelse{\boolean{articletitles}}{\emph{{The EvtGen particle
  decay simulation package}},
  }{}\href{http://dx.doi.org/10.1016/S0168-9002(01)00089-4}{Nucl.\ Instrum.\
  Meth.\  \textbf{A462} (2001) 152}\relax
\mciteBstWouldAddEndPuncttrue
\mciteSetBstMidEndSepPunct{\mcitedefaultmidpunct}
{\mcitedefaultendpunct}{\mcitedefaultseppunct}\relax
\EndOfBibitem
\bibitem{Agostinelli:2002hh}
Geant4 collaboration, S.~Agostinelli {\em et~al.},
  \ifthenelse{\boolean{articletitles}}{\emph{{Geant4: A simulation toolkit}},
  }{}\href{http://dx.doi.org/10.1016/S0168-9002(03)01368-8}{Nucl.\ Instrum.\
  Meth.\  \textbf{A506} (2003) 250}\relax
\mciteBstWouldAddEndPuncttrue
\mciteSetBstMidEndSepPunct{\mcitedefaultmidpunct}
{\mcitedefaultendpunct}{\mcitedefaultseppunct}\relax
\EndOfBibitem
\bibitem{Allison:2006ve}
Geant4 collaboration, J.~Allison {\em et~al.},
  \ifthenelse{\boolean{articletitles}}{\emph{{Geant4 developments and
  applications}}, }{}\href{http://dx.doi.org/10.1109/TNS.2006.869826}{IEEE
  Trans.\ Nucl.\ Sci.\  \textbf{53} (2006) 270}\relax
\mciteBstWouldAddEndPuncttrue
\mciteSetBstMidEndSepPunct{\mcitedefaultmidpunct}
{\mcitedefaultendpunct}{\mcitedefaultseppunct}\relax
\EndOfBibitem
\bibitem{Golonka:2005pn}
P.~Golonka and Z.~Was, \ifthenelse{\boolean{articletitles}}{\emph{{PHOTOS Monte
  Carlo: A precision tool for QED corrections in $Z$ and $W$ decays}},
  }{}\href{http://dx.doi.org/10.1140/epjc/s2005-02396-4}{Eur.\ Phys.\ J.\
  \textbf{C45} (2006) 97},
  \href{http://arxiv.org/abs/hep-ph/0506026}{{\normalfont\ttfamily
  arXiv:hep-ph/0506026}}\relax
\mciteBstWouldAddEndPuncttrue
\mciteSetBstMidEndSepPunct{\mcitedefaultmidpunct}
{\mcitedefaultendpunct}{\mcitedefaultseppunct}\relax
\EndOfBibitem
\bibitem{LHCb-PROC-2011-006}
M.~Clemencic {\em et~al.}, \ifthenelse{\boolean{articletitles}}{\emph{{The
  \lhcb simulation application, Gauss: Design, evolution and experience}},
  }{}\href{http://dx.doi.org/10.1088/1742-6596/331/3/032023}{{J.\ Phys.\ Conf.\
  Ser.\ } \textbf{331} (2011) 032023}\relax
\mciteBstWouldAddEndPuncttrue
\mciteSetBstMidEndSepPunct{\mcitedefaultmidpunct}
{\mcitedefaultendpunct}{\mcitedefaultseppunct}\relax
\EndOfBibitem
\bibitem{Feindt:2006pm}
M.~Feindt and U.~Kerzel, \ifthenelse{\boolean{articletitles}}{\emph{{The
  NeuroBayes neural network package}},
  }{}\href{http://dx.doi.org/10.1016/j.nima.2005.11.166}{Nucl.\ Instrum.\
  Meth.\  \textbf{A559} (2006) 190}\relax
\mciteBstWouldAddEndPuncttrue
\mciteSetBstMidEndSepPunct{\mcitedefaultmidpunct}
{\mcitedefaultendpunct}{\mcitedefaultseppunct}\relax
\EndOfBibitem
\bibitem{LHCb-PAPER-2013-046}
LHCb collaboration, R.~Aaij {\em et~al.},
  \ifthenelse{\boolean{articletitles}}{\emph{{Measurement of the
  $B_s^0\to\mu^+\mu^-$ branching fraction and search for $B^0\to\mu^+\mu^-$
  decays at the LHCb experiment}},
  }{}\href{http://dx.doi.org/10.1103/PhysRevLett.111.101805}{Phys.\ Rev.\
  Lett.\  \textbf{111} (2013) 101805},
  \href{http://arxiv.org/abs/1307.5024}{{\normalfont\ttfamily
  arXiv:1307.5024}}\relax
\mciteBstWouldAddEndPuncttrue
\mciteSetBstMidEndSepPunct{\mcitedefaultmidpunct}
{\mcitedefaultendpunct}{\mcitedefaultseppunct}\relax
\EndOfBibitem
\bibitem{LHCb-PAPER-2014-052}
LHCb collaboration, R.~Aaij {\em et~al.},
  \ifthenelse{\boolean{articletitles}}{\emph{{Search for the lepton flavour
  violating decay $\tau^-\to \mu^-\mu^+\mu^-$}},
  }{}\href{http://dx.doi.org/10.1007/JHEP10(2015)121}{JHEP \textbf{02} (2015)
  121}, \href{http://arxiv.org/abs/1409.8548}{{\normalfont\ttfamily
  arXiv:1409.8548}}\relax
\mciteBstWouldAddEndPuncttrue
\mciteSetBstMidEndSepPunct{\mcitedefaultmidpunct}
{\mcitedefaultendpunct}{\mcitedefaultseppunct}\relax
\EndOfBibitem
\bibitem{LHCb-PAPER-2015-012}
LHCb collaboration, R.~Aaij {\em et~al.},
  \ifthenelse{\boolean{articletitles}}{\emph{{Search for the decay
  $B^0_s\to\overline{D}^0f_0(980)$}},
  }{}\href{http://dx.doi.org/10.1007/JHEP08(2015)005}{JHEP \textbf{08} (2015)
  005}, \href{http://arxiv.org/abs/1505.01654}{{\normalfont\ttfamily
  arXiv:1505.01654}}\relax
\mciteBstWouldAddEndPuncttrue
\mciteSetBstMidEndSepPunct{\mcitedefaultmidpunct}
{\mcitedefaultendpunct}{\mcitedefaultseppunct}\relax
\EndOfBibitem
\bibitem{LHCb-PAPER-2015-048}
LHCb collaboration, R.~Aaij {\em et~al.},
  \ifthenelse{\boolean{articletitles}}{\emph{{Search for the lepton-flavour
  violating decay $D^0 \to e^\pm\mu^\mp$}},
  }{}\href{http://arxiv.org/abs/1512.00322}{{\normalfont\ttfamily
  arXiv:1512.00322}}, {submitted to Phys.Lett.B}\relax
\mciteBstWouldAddEndPuncttrue
\mciteSetBstMidEndSepPunct{\mcitedefaultmidpunct}
{\mcitedefaultendpunct}{\mcitedefaultseppunct}\relax
\EndOfBibitem
\bibitem{Skwarnicki:1986xj}
T.~Skwarnicki, {\em {A study of the radiative cascade transitions between the
  Upsilon-prime and Upsilon resonances}}, PhD thesis, Institute of Nuclear
  Physics, Krakow, 1986,
  {\href{http://inspirehep.net/record/230779/}{DESY-F31-86-02}}\relax
\mciteBstWouldAddEndPuncttrue
\mciteSetBstMidEndSepPunct{\mcitedefaultmidpunct}
{\mcitedefaultendpunct}{\mcitedefaultseppunct}\relax
\EndOfBibitem
\bibitem{Abe:2003zm}
Belle collaboration, K.~Abe {\em et~al.},
  \ifthenelse{\boolean{articletitles}}{\emph{{Study of $\Bm \to D^{**0} \pim \
  (D^{**0} \to D^{(*)+} \pim)$ decays}},
  }{}\href{http://dx.doi.org/10.1103/PhysRevD.69.112002}{Phys.\ Rev.\
  \textbf{D69} (2004) 112002},
  \href{http://arxiv.org/abs/hep-ex/0307021}{{\normalfont\ttfamily
  arXiv:hep-ex/0307021}}\relax
\mciteBstWouldAddEndPuncttrue
\mciteSetBstMidEndSepPunct{\mcitedefaultmidpunct}
{\mcitedefaultendpunct}{\mcitedefaultseppunct}\relax
\EndOfBibitem
\bibitem{Aubert:2009wg}
\babar collaboration, B.~Aubert {\em et~al.},
  \ifthenelse{\boolean{articletitles}}{\emph{{Dalitz plot analysis of $\Bm \to
  \Dp\pim\pim$}}, }{}\href{http://dx.doi.org/10.1103/PhysRevD.79.112004}{Phys.\
  Rev.\  \textbf{D79} (2009) 112004},
  \href{http://arxiv.org/abs/0901.1291}{{\normalfont\ttfamily
  arXiv:0901.1291}}\relax
\mciteBstWouldAddEndPuncttrue
\mciteSetBstMidEndSepPunct{\mcitedefaultmidpunct}
{\mcitedefaultendpunct}{\mcitedefaultseppunct}\relax
\EndOfBibitem
\bibitem{PDG2014}
Particle Data Group, K.~A. Olive {\em et~al.},
  \ifthenelse{\boolean{articletitles}}{\emph{{\href{http://pdg.lbl.gov/}{Review
  of particle physics}}},
  }{}\href{http://dx.doi.org/10.1088/1674-1137/38/9/090001}{Chin.\ Phys.\
  \textbf{C38} (2014) 090001}\relax
\mciteBstWouldAddEndPuncttrue
\mciteSetBstMidEndSepPunct{\mcitedefaultmidpunct}
{\mcitedefaultendpunct}{\mcitedefaultseppunct}\relax
\EndOfBibitem
\bibitem{LHCb-PAPER-2012-033}
LHCb collaboration, R.~Aaij {\em et~al.},
  \ifthenelse{\boolean{articletitles}}{\emph{{First observation of the decays
  $\overline{B}^0_{(s)} \to D_s^+ K^- \pi^+\pi^-$ and $\overline{B}^0_s \to
  D_{s1}(2536)^+\pi^-$}},
  }{}\href{http://dx.doi.org/10.1103/PhysRevD.86.112005}{Phys.\ Rev.\
  \textbf{D86} (2012) 112005},
  \href{http://arxiv.org/abs/1211.1541}{{\normalfont\ttfamily
  arXiv:1211.1541}}\relax
\mciteBstWouldAddEndPuncttrue
\mciteSetBstMidEndSepPunct{\mcitedefaultmidpunct}
{\mcitedefaultendpunct}{\mcitedefaultseppunct}\relax
\EndOfBibitem
\bibitem{LHCb-PAPER-2014-035}
LHCb collaboration, R.~Aaij {\em et~al.},
  \ifthenelse{\boolean{articletitles}}{\emph{{Observation of overlapping
  spin-$1$ and spin-$3$ $\overline{D}^0 K^-$ resonances at mass
  $2.86$~GeV/$c^2$}},
  }{}\href{http://dx.doi.org/10.1103/PhysRevLett.113.162001}{Phys.\ Rev.\
  Lett.\  \textbf{113} (2014) 162001},
  \href{http://arxiv.org/abs/1407.7574}{{\normalfont\ttfamily
  arXiv:1407.7574}}\relax
\mciteBstWouldAddEndPuncttrue
\mciteSetBstMidEndSepPunct{\mcitedefaultmidpunct}
{\mcitedefaultendpunct}{\mcitedefaultseppunct}\relax
\EndOfBibitem
\bibitem{LHCb-PAPER-2014-036}
LHCb collaboration, R.~Aaij {\em et~al.},
  \ifthenelse{\boolean{articletitles}}{\emph{{Dalitz plot analysis of
  $B^0_s\to\overline{D}^0K^-\pi^+$ decays}},
  }{}\href{http://dx.doi.org/10.1103/PhysRevD.90.072003}{Phys.\ Rev.\
  \textbf{D90} (2014) 072003},
  \href{http://arxiv.org/abs/1407.7712}{{\normalfont\ttfamily
  arXiv:1407.7712}}\relax
\mciteBstWouldAddEndPuncttrue
\mciteSetBstMidEndSepPunct{\mcitedefaultmidpunct}
{\mcitedefaultendpunct}{\mcitedefaultseppunct}\relax
\EndOfBibitem
\bibitem{Pivk:2004ty}
M.~Pivk and F.~R. Le~Diberder,
  \ifthenelse{\boolean{articletitles}}{\emph{{sPlot: A statistical tool to
  unfold data distributions}},
  }{}\href{http://dx.doi.org/10.1016/j.nima.2005.08.106}{Nucl.\ Instrum.\
  Meth.\  \textbf{A555} (2005) 356},
  \href{http://arxiv.org/abs/physics/0402083}{{\normalfont\ttfamily
  arXiv:physics/0402083}}\relax
\mciteBstWouldAddEndPuncttrue
\mciteSetBstMidEndSepPunct{\mcitedefaultmidpunct}
{\mcitedefaultendpunct}{\mcitedefaultseppunct}\relax
\EndOfBibitem
\bibitem{LHCb-DP-2012-003}
M.~Adinolfi {\em et~al.},
  \ifthenelse{\boolean{articletitles}}{\emph{{Performance of the \lhcb RICH
  detector at the LHC}},
  }{}\href{http://dx.doi.org/10.1140/epjc/s10052-013-2431-9}{Eur.\ Phys.\ J.\
  \textbf{C73} (2013) 2431},
  \href{http://arxiv.org/abs/1211.6759}{{\normalfont\ttfamily
  arXiv:1211.6759}}\relax
\mciteBstWouldAddEndPuncttrue
\mciteSetBstMidEndSepPunct{\mcitedefaultmidpunct}
{\mcitedefaultendpunct}{\mcitedefaultseppunct}\relax
\EndOfBibitem
\end{mcitethebibliography}

%\clearpage
%\input{supplementary-app}

\clearpage
\centerline{\large\bf LHCb collaboration}
\begin{flushleft}
\small
R.~Aaij$^{39}$, 
C.~Abell\'{a}n~Beteta$^{41}$, 
B.~Adeva$^{38}$, 
M.~Adinolfi$^{47}$, 
A.~Affolder$^{53}$, 
Z.~Ajaltouni$^{5}$, 
S.~Akar$^{6}$, 
J.~Albrecht$^{10}$, 
F.~Alessio$^{39}$, 
M.~Alexander$^{52}$
% ...

{\footnotesize \it
% ...
$ ^{5}$Clermont Universit\'{e}, Universit\'{e} Blaise Pascal, CNRS/IN2P3, LPC, Clermont-Ferrand, France\\
$ ^{6}$CPPM, Aix-Marseille Universit\'{e}, CNRS/IN2P3, Marseille, France\\
% ...
$ ^{10}$Fakult\"{a}t Physik, Technische Universit\"{a}t Dortmund, Dortmund, Germany\\
% ...
$ ^{38}$Universidad de Santiago de Compostela, Santiago de Compostela, Spain\\
$ ^{39}$European Organization for Nuclear Research (CERN), Geneva, Switzerland\\
% ...
$ ^{41}$Physik-Institut, Universit\"{a}t Z\"{u}rich, Z\"{u}rich, Switzerland\\
% ...
$ ^{47}$H.H. Wills Physics Laboratory, University of Bristol, Bristol, United Kingdom\\
% ...
$ ^{52}$School of Physics and Astronomy, University of Glasgow, Glasgow, United Kingdom\\
$ ^{53}$Oliver Lodge Laboratory, University of Liverpool, Liverpool, United Kingdom\\
% ...
}
\end{flushleft}

\end{document}